\documentclass[aps,prd,showpacs,superscriptaddress,nofootinbib,twocolumn]{revtex4}
\usepackage{graphicx,color}
\usepackage{amsmath}
\usepackage[sort&compress]{natbib}
\usepackage[normalem]{ulem}

\usepackage[export]{adjustbox}

\begin{document}
\title{Improved Limits on Sterile Neutrino Dark Matter using Full-Sky Fermi Gamma-Ray Burst Monitor Data}

\author{Kenny C.~Y.~Ng}
\email{ng.199@osu.edu}
\affiliation{Center for Cosmology and AstroParticle Physics (CCAPP), Ohio State University, Columbus, OH 43210}
\affiliation{Department of Physics, Ohio State University, Columbus, OH 43210}

\author{Shunsaku Horiuchi}
\email{horiuchi@vt.edu}
\affiliation{Center for Cosmology, Department of Physics and Astronomy, University of California, Irvine, CA 92697, USA}
\affiliation{Center for Neutrino Physics, Department of Physics, Virginia Tech, 
Blacksburg, VA 24061, USA}

\author{Jennifer M.~Gaskins}
\email{jgaskins@uva.nl}
\affiliation{California Institute of Technology, Pasadena, CA 91125, USA}
\affiliation{GRAPPA, University of Amsterdam, 1098 XH Amsterdam, Netherlands}

\author{Miles Smith}
\email{miles.smith@jpl.nasa.gov}
\affiliation{Pennsylvania State University, PA 16802}

\author{Robert Preece}
\email{rob.preece@nasa.gov}
\affiliation{Department of Space Science, University of Alabama in Huntsville, Huntsville, AL 35899}

\date{May 21, 2015}

\begin{abstract}
A sterile neutrino of $\sim$\,keV mass is a well motivated dark matter candidate.  Its decay generates an X-ray line that offers a unique target for X-ray telescopes.  For the first time, we use the Gamma-ray Burst Monitor (GBM) onboard the {\it Fermi} Gamma-Ray Space Telescope to search for sterile neutrino decay lines; our analysis covers the energy range 10--25\,keV\,(sterile neutrino mass 20--50\,keV), which is  inaccessible to X-ray and gamma-ray satellites such as {\it Chandra}, {\it Suzaku}, {\it XMM-Newton}, and {\it INTEGRAL}. The extremely wide field of view of the GBM enables a large fraction of the Milky Way dark matter halo to be probed.  After implementing careful data cuts, we obtain $\sim$\,53 days of full sky observational data.  We observe an excess of photons towards the Galactic Center, as expected from astrophysical emission.  We search for sterile neutrino decay lines in the energy spectrum, and find no significant signal.  From this, we obtain upper limits on the sterile neutrino mixing angle as a function of mass.  In the sterile neutrino mass range 25--40\,keV, we improve upon previous upper limits by approximately an order of magnitude.  Better understanding of detector and astrophysical backgrounds, as well as detector response, will further improve the sensitivity of a search with the GBM\@.

\end{abstract}



\maketitle

\section{Introduction}

Right-handed neutral fermions (henceforth sterile neutrinos) arise in many extensions of the Standard Model in explaining the observed flavor oscillations of active neutrinos, and yield an extremely rich phenomenology (for recent reviews, see, e.g., \cite{Abazajian:2012ys}). 
Sterile neutrinos may  be produced in core-collapse supernovae \cite{Kusenko:1997sp}, providing a new mechanism for explosion \cite{Hidaka:2006sg}, and may explain the origin of strong neutron star kicks \cite{Fuller:2003gy,Barkovich:2004jp,Fryer:2005sz}.  The sterile neutrino can modify big bang nucleosynthesis \cite{Dolgov:2000jw,Abazajian:2004aj,Smith:2008ic}, assist reionization \cite{Barkana:2001gr,Hansen:2003yj, Mapelli:2005hq, Biermann:2006bu,Mapelli:2006ej, Yeung:2012ya}, and affect neutrino oscillations \cite{Cirelli:2004cz}. 

Moreover, it has been noted that sterile neutrinos can contribute the entirety of the observed dark matter density. They could be produced in the early universe via oscillation mechanisms, including non-resonantly~\cite{Dodelson:1993je} or resonantly with active neutrinos~\cite{Shi:1998km}, or alternatively via non-oscillation mechanisms, such as decays of heavy particles~(see \cite{Shaposhnikov:2006xi, Kusenko:2006rh, Merle:2013wta, Frigerio:2014ifa, Lello:2014yha, Merle:2015oja} for some of the scenarios).  Sterile neutrinos could be a warm or cold dark matter candidate~\cite{Abazajian:2001nj,Asaka:2006nq,Petraki:2007gq, Boyarsky:2008xj,Boyanovsky:2010pw}.  In addition, some sterile neutrino dark matter models can explain the baryon asymmetry in the Universe \cite{Akhmedov:1998qx,Asaka:2005an,Asaka:2005pn,Boyarsky:2009ix}.

For sterile neutrinos produced via oscillations to be a viable dark matter candidate, they typically have mass in the 1 -- 100\,keV range \cite{Dodelson:1993je,Shi:1998km,Abazajian:2001nj}.  
They can radiatively decay into an active neutrino and a photon \cite{Pal:1981rm,Barger:1995ty}.  
The photon carries half of the total energy, and therefore lies in the X-ray energy range.  
The photon line is strongly distinct from most astrophysical and detector backgrounds, which have smooth energy spectra.  An exception is line emissions from hot gases and activated detector materials. 
While the decay lifetime must be comparable to the age of the Universe in order to ensure that sterile neutrinos remain a viable dark matter candidate, the decay in high concentrations of dark matter, e.g., centers of galaxies, clusters of galaxies, and dwarf spheroidal galaxies, can lead to an appreciable X-ray flux.  
The large expected flux in many targets, coupled with the spectral and morphological characteristics of the signal, make searches with X-rays a very powerful approach for testing sterile neutrino dark matter scenarios~(see \cite{Kusenko:2009up,Boyarsky:2009ix, Kusenko:2013saa} for a comprehensive discussion of various searches). 

The X-ray constraint on sterile neutrino dark matter was first obtained using the Cosmic X-ray Background~(CXB)~\cite{Dolgov:2000ew}.  Subsequently a more detailed analysis was carried out in \cite{Abazajian:2001vt}, where the authors considered a set of galaxy clusters, two spiral galaxies, and the Cosmic X-ray Background (CXB), and used observations by {\it Chandra} and {\it XMM-Newton}.  Since then, a host of other sources have been explored~(for a full discussion, see, e.g., review articles~\cite{Kusenko:2009up,Boyarsky:2009ix}), including other galaxy clusters such as Coma \cite{Boyarsky:2006zi,Abazajian:2006yn}, the distant A$520$ \cite{RiemerSorensen:2006pi}, and the Bullet \cite{Boyarsky:2006kc}; nearby galaxies such as Andromeda \cite{Watson:2006qb,Boyarsky:2007ay,Watson:2011dw,Horiuchi:2013noa} and M33 \cite{Borriello:2011un}; additional analysis of the CXB \cite{Boyarsky:2005us,Boyarsky:2006fg,Abazajian:2006jc,  Sekiya:2015jsa}; more recently the Milky Way satellites including the Large Magellanic Cloud \cite{Boyarsky:2006fg}, Ursa Minor \cite{Boyarsky:2006ag,Loewenstein:2008yi}, Draco \cite{RiemerSorensen:2009jp}, Willman I \cite{Loewenstein:2009cm}, and Segue I \cite{Mirabal:2010an}; and, finally, the nearest dark matter concentration, the Milky Way galaxy \cite{Boyarsky:2006fg,Boyarsky:2006ag,RiemerSorensen:2006fh,Abazajian:2006jc,Yuksel:2007xh,Boyarsky:2007ge,Prokhorov:2010us}. 

Several possible line detections consistent with sterile neutrino dark matter decay have been reported.
A line feature was observed in Willman I that could be interpreted as the decay of a sterile neutrino of mass $m_s \approx 5$ keV and sin$^2 \theta \approx 10^{-9}$, where $\theta$ is the mixing angle between the sterile and active neutrinos \cite{Loewenstein:2009cm}~(but, see~\cite{2010MNRAS.407.1188B, 2012ApJ...751...82L}). In another study, X-ray line ratios showed an excess that could be interpreted as arising from the decays of 17\,keV sterile neutrinos with sin$^2 \theta \approx 10^{-12}$ \cite{Prokhorov:2010us}.  Most recently, an anomolous X-ray line was detected from galaxy clusters and Andromeda \cite{Bulbul:2014sua, Boyarsky:2014jta} (also see \cite{Riemer-Sorensen:2014yda, Jeltema:2014qfa, Boyarsky:2014ska, Malyshev:2014xqa, Anderson:2014tza, Carlson:2014lla, 2015PASJ...67...23T, Bonnivard:2015gla}), which can be interpreted as the decay of 7\,keV sterile neutrinos \cite{Abazajian:2014gza}. 

In this work, we use the Gamma-ray Burst Monitor (GBM) onboard the \emph{Fermi} Gamma-Ray Space Telescope to search for X-ray lines.  Notable advantages of the GBM include its all-sky coverage, which allows the entire Milky Way dark matter halo to be explored, and large effective area, yielding a very high statistics data set. The energy range of the GBM extends from 8\,keV up to 40\,MeV, conveniently filling a gap in energy above the range of previously considered X-ray satellites and below the range of {\it INTEGRAL}.  Therefore, in this work we focus on this unexplored photon energy range $E_{\gamma}$ = 10--25\,keV\,($m_{s}$ = 20--50\,keV).   
We consider the Milky Way because of its proximity and well-studied dark matter distribution, and because the GBM detectors are more sensitive to large scale diffuse emission such as from the Galactic halo, due to GBM's large field of view~(FOV) and poor angular resolution.

We describe the expected X-ray signal from sterile neutrino dark matter decays in Sec.~\ref{sec:theory}. The GBM instrument and the dark matter signal modeling in the context of the GBM detectors are presented in Sec.~\ref{sec:instrument}.  The data reduction procedures are described in Sec.~\ref{sec:observations}.  In Sec.~\ref{sec:limits}, we describe the line search analysis and the procedure used to obtain limits on sterile neutrino decays.  We summarize in Sec.~\ref{sec:disconclu}.  Throughout this work, we adopt cosmological parameters from Planck \cite{Ade:2013zuv}, where $H_{0} = 100 h~{\rm km/s/Mpc}$, $h=0.673$, $\Omega_{\Lambda} = 0.685$, $\Omega_{M} = 0.315$, $h(z) = \sqrt{\Omega_{\Lambda} + \Omega_{M}(1+z)^{3}}$, the dark matter fraction $\Omega_{DM} = 0.265$, and $\rho_{c} = 1.05\times 10^{-5}h^{2}~{\rm GeV~cm^{-3}}$.

\section{Expected signal flux \label{sec:theory}}

\begin{figure}[t]
\includegraphics[width=3.25in]{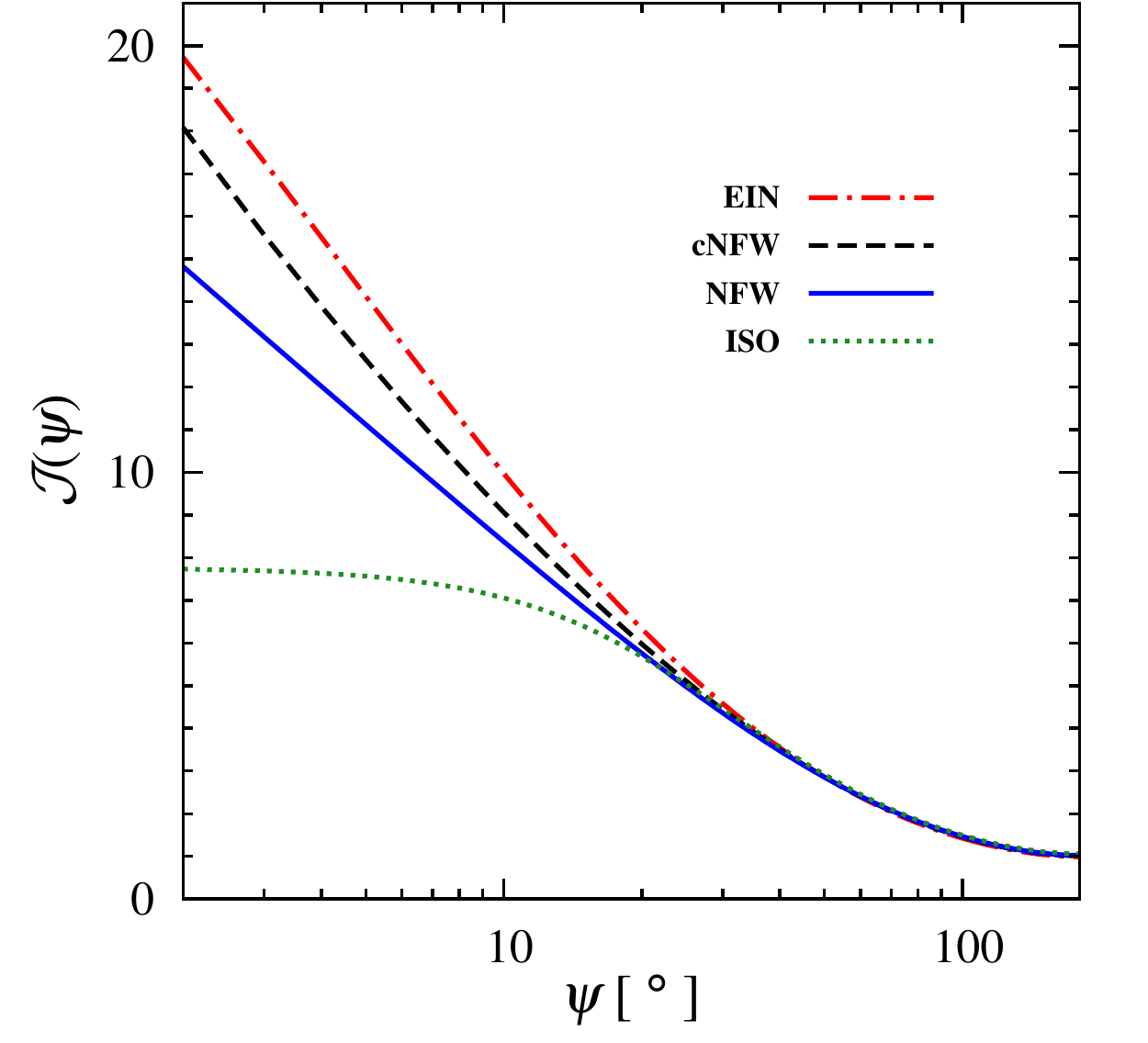}
\caption{\label{fig:Jfactor} The J-factor, $ {\cal J}(\psi)$~(Eq.~(\ref{eq:jfactor})), as a function of half opening angle $\psi$ relative to the GC, for four Milky Way dark matter halo profiles. }
\end{figure}

The primary decay channel of sterile neutrinos is into three light active neutrinos. The radiative decay into an active neutrino and a photon that we are interested in is suppressed by a factor of $27 \alpha / 8 \pi \approx 1/128$ relative to the primary decay channel \cite{Barger:1995ty}, and has a decay rate \citep{Pal:1981rm,Abazajian:2001vt}
\begin{equation}\label{eq:countrate}
\Gamma_s \simeq 1.36 \times 10^{-32} \, {\rm s^{-1}} \left( \frac{{\rm sin}^2 2\theta}{10^{-10}} \right) \left( \frac{m_s}{\rm 1 \, keV} \right)^5\,,
\end{equation}
where we have assumed a Majorana sterile neutrino (for a Dirac sterile neutrino the decay rate is halved). The energy luminosity of decay photons arising from a sterile neutrino dark matter clump of mass $M_{\rm DM}$ is given by $L_\gamma = E_\gamma (M_{\rm DM} / m_s) \Gamma_s$, where $E_\gamma = m_s / 2$ is the photon energy, and equals
\begin{equation}
L_\gamma \simeq 1.2 \times 10^{38} \, {\rm erg \, s^{-1}}  \left( \frac{M_{\rm DM}}{10^{11} M_\odot} \right)  \left( \frac{{\rm sin}^2 2\theta}{10^{-10}} \right) \left( \frac{m_s}{\rm 10 \, keV} \right)^5\,,
\end{equation}
for a typical galaxy-size dark matter halo mass. It can be immediately appreciated that this is comparable to the total luminosity of astrophysical X-rays in the Milky Way in the $2-10$ keV range, $\sim 10^{39} \, {\rm erg \, s^{-1}}$ \citep{Grimm:2001vd}, or the total Milky Way diffuse emission in the same energy range, $\sim 10^{38} \, {\rm erg \, s^{-1}}$ \citep{Dogiel:2001fg}.

\begin{table}
\caption{\label{tab:table1}Dark matter profile parameters for widely adopted dark matter profiles in the literature. Our canonical profile is the NFW profile.}
\begin{ruledtabular}
\begin{tabular}{lccccc}
\hline
Profile		& $\alpha$	& $\beta$ 		& $\gamma$	& $R_s$ [kpc]		\\
\hline
NFW			& $1$		& $3$ 		& $1$		& $20$			 	\\	
cNFW		& $1$		& $3$ 		& $1.15$		& $23.7$			 	\\
Cored isothermal~(ISO)	& $2$		& $2$ 		& $0$		& $3.5$		 		\\	
Einasto~(EIN)		& -			& - 			& -			& $20$			 	\\	
\end{tabular}
\end{ruledtabular}
\end{table}

The photon intensity (number flux per solid angle) of sterile neutrino dark matter decay coming from an angle $\psi$ away from the Galactic Center~(GC) consists of both the Galactic and the extragalactic components, 
\begin{eqnarray}\label{eq:Intensity}
{\cal I}(\psi,E) &\equiv& \frac{dN}{dAdTd\Omega dE} \\
&=&  \frac{\rho_\odot R_\odot} {4\pi m_s \tau_s} 
 {\cal J}(\psi)\frac{dN}{dE} +  \frac{\Omega_{DM} \rho_{c}} {4\pi m_s \tau_s} \frac{c}{H_{0}} \int \frac{ dz}{h(z)}\frac{dN}{dE^{\prime}}  \nonumber \, \\
 &=&  \frac{\rho_\odot R_\odot} {4\pi m_s \tau_s} \left(
 {\cal J}(\psi)\frac{dN}{dE} + R_{\rm EG} \int \frac{ dz}{h(z)}\frac{dN}{dE^{\prime}} \right) \nonumber \,,
\end{eqnarray}
where $\tau_{s} = 1/\Gamma_s$ is the lifetime, $\rho_\odot = 0.3 \, {\rm GeV \, cm^{-3}}$ is the local dark matter mass density, $R_\odot = 8.5$\,kpc is the Sun's distance to the GC, and $dN/dE = \delta(E - m_{s}/2)$ is the dark matter decay spectrum. The first term in the bracket is the Galactic component.  The so-called J-factor, ${\cal J}(\psi)$, is the integral of the dark matter mass density $\rho$ in the Milky Way halo along the line-of-sight,
\begin{equation}\label{eq:jfactor}
{\cal J}(\psi) = \frac{1}{\rho_\odot R_\odot} \int_{0}^{\ell_{max}} d\ell \; 
\rho(\psi,\ell) \, , 
\end{equation}
where $\ell_{max}$ is the outer limit of the dark matter halo.  We assume the dark matter distribution is spherically symmetric about the GC, hence
\begin{equation}
\rho(\psi,\ell)=\rho(r_{\rm GC}(\psi,\ell))=\rho\left(\sqrt{R_\odot^2-2\, \ell\, R_\odot\cos\psi+\ell^2} \right)\, .
\end{equation}
The value of $\ell_{max}$ differs depending on the adopted halo model, but the contribution to ${\cal J}(\psi) $ from beyond $\sim 30$\,kpc is negligible. We adopt $\ell_{max}=250$\,kpc in this work.   

The second term in the bracket of Eq.~(\ref{eq:Intensity}) describes the isotropic extragalactic component, where $E^{\prime} = E(1+z)$.  The factor $R_{\rm EG}$ roughly compares the contribution of the extragalactic component versus the Galactic component, up to the shape of the energy spectrum.
\begin{equation}
R_{\rm EG} \equiv \frac{c}{H_{0}}\frac{\Omega_{DM} \rho_{c}}{\rho_\odot R_\odot} \simeq 2\, .
\end{equation}
Normally, the extragalactic component can be ignored as typically the analysis region is chosen to be a small patch of the sky where the Galactic component is much larger~(e.g., the GC, where ${\cal J}\gg 1$).  However, in our case, the large FOV of the GBM makes the extragalactic component non-negligible.  

The dark matter density profile $\rho(r)$ of the Milky Way is not precisely known, in particular at small Galactic radius.  We consider several fitting functions that capture the results of numerical simulations of dark matter halo profiles, which can be parameterized by the following form,
\begin{equation}
\rho^{\alpha\beta\gamma} (r) = \rho_\odot \left( \frac{r}{R_\odot} \right)^{-\gamma} \left[ \frac{1+(R_\odot/R_s)^\alpha}{1+(r/R_s)^\alpha} \right]^{(\beta-\gamma)/\alpha},
\end{equation}
where parameters for commonly used profiles are summarized in Table \ref{tab:table1}. Another profile favored by recent simulations is the Einasto profile, 
\begin{equation}
\rho^{\rm Ein} (r) = \rho_\odot \, {\rm exp}\left( -\frac{2}{\alpha_E}\frac{r^{\alpha_E}-R_\odot^{\alpha_E}}{R_s^{\alpha_E}} \right),
\end{equation}
with $\alpha_E = 0.17$ and scale radius $R_s = 20$\,kpc. 
These profiles differ mainly at small Galactic radius.  The first three profiles have constant logarithmic slopes at small radii, which are described by the $\gamma$ factor.  The Einasto profile has the same slope as the NFW profile at the scale radius, but the slope decreases as the radius decreases.

In Fig.~\ref{fig:Jfactor}, we show the J-factor ${\cal J}(\psi)$ for each dark matter profile as a function of the angle $\psi$ viewed away from the GC. The differences between profiles are relatively small, because the density $\rho$ appears linearly in the decay flux (as opposed to in the annihilation flux where the density appears quadratically).  We use the NFW profile as our canonical profile in this work.  As will be shown in Sec.~\ref{sec:modeling}, the impact of varying the profile is minimal after taking into account the detector response and the FOV.  Thus the sterile neutrino constraint obtained using GBM is robust against dark matter profile uncertainties.  

A crude estimate of the expected number of photons $\nu_{\gamma}$ per unit time $T$ from Galactic dark matter decay is 
\begin{eqnarray}
\frac{d\nu_\gamma}{dT} &\sim& 20\,{\rm s}^{-1}  \left( \frac{A_{\rm eff}\Omega}{20\pi \, {\rm cm^2\,sr}} \right) \left( \frac{{\cal J}_{60}}{2} \right) \times \nonumber\\
 && \left( \frac{{\rm sin}^2 2\theta}{10^{-11}} \right) \left( \frac{m_s}{\rm 20 \, keV} \right)^4,  
\end{eqnarray}
where we use representative values for the effective area and solid angle, the J-factor at $\psi = 60^{\circ}$, ${\cal{J}}_{60}$, and a nominal sterile neutrino mixing angle. It is immediately clear that even a small fraction of the total Fermi-GBM live time can yield significant number of signal photons. 
\section{Instrument and signal modeling \label{sec:instrument}}

\subsection{GBM Instrumentation}
The GBM consists of 14 detectors: 12 NaI detectors, each operating over energies from 8\,keV to 1\,MeV, and 2 BGO detectors, each operating over energies from 200\,keV to 40\,MeV. The NaI detectors are located on the corners and sides of the spacecraft, with different orientations, and they together provide a nearly complete coverage of the occulted sky. At any given time, typically 3--4 NaI detectors view the Earth within 60 degrees of the detector zenith, i.e., their FOV is occulted by the Earth. 

\begin{figure}[t]
\includegraphics[width=3.25in]{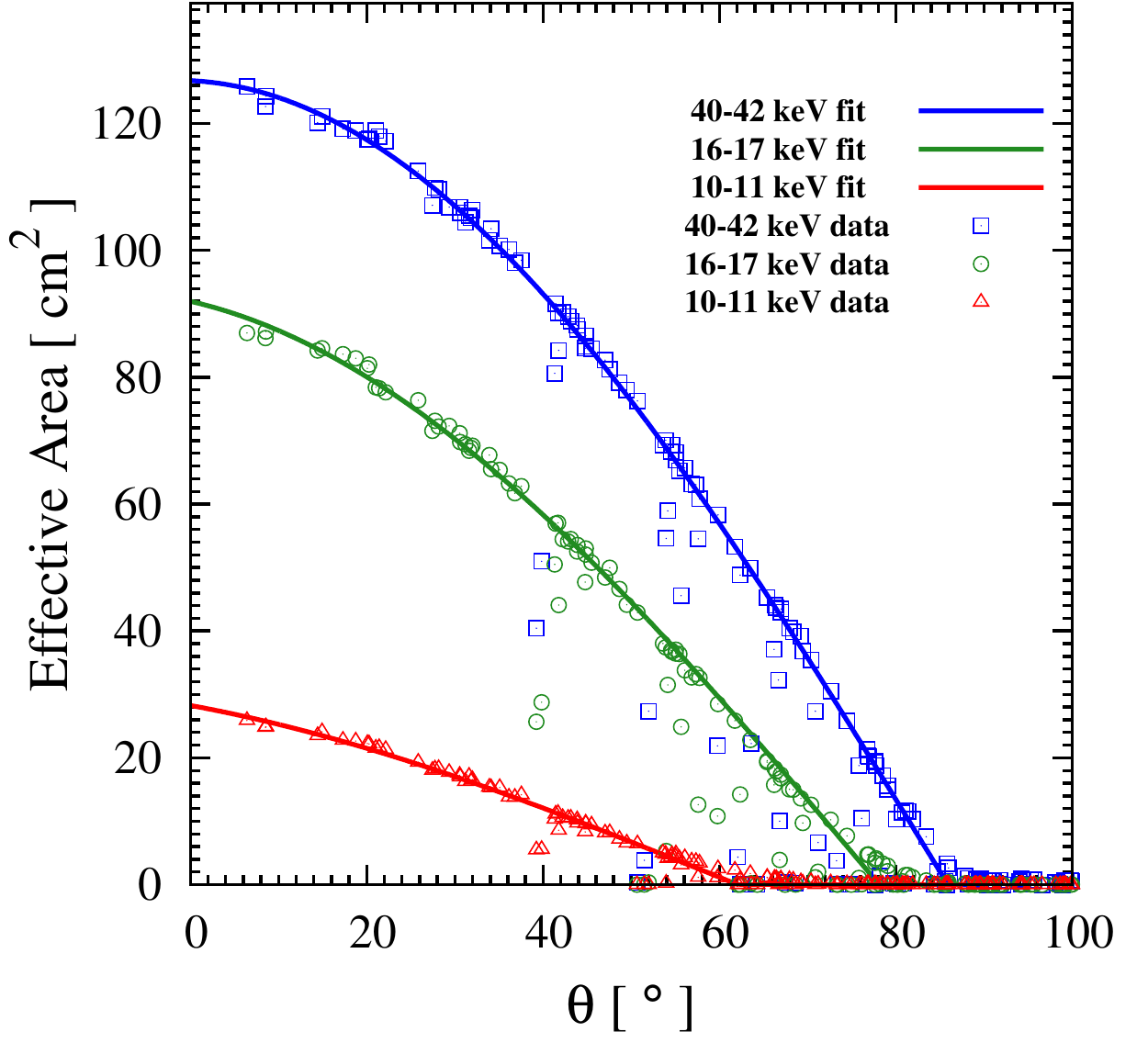} 
\caption{The effective area for det-7 NaI detector versus the detector zenith angle relative to the detector normal, for three example energy bins.  Points are data from GBM calibration files, and the anomalous dips in the effective area come from blockages from other satellite components.  Solid lines are fits to the data neglecting the dips. \label{fig:effa_angle}}
\end{figure}

Not all of the NaI detectors are best suited for dark matter searches.  At first consideration, det-0 and det-6 would seem to be the best detectors to use since they are aligned close to the LAT zenith~($\simeq 20^{\circ}$ offset).  However, we find that significant parts of the FOV of these two detectors are actually blocked by the LAT itself.  Also, half of the detectors are pointed towards the Sun all the time, and X-ray emissions from the Sun contaminate their low energy spectrum.  Lastly, some detectors are pointed sideways, i.e. $\simeq 90^{\circ}$ relative the LAT-zenith, which suffer large FOV blockage from the Earth.  Ruling out these detectors, only det-7 and det-9 seem to be suitable, which are $\simeq 45^{\circ}$ relative the LAT-zenith.  Upon inspection, we observe an anomalous spectral feature in the low energy spectrum of det-9 compared to other detectors.  As a result, we use det-7 as our fiducial detector for analysis.  
As will be shown below, this analysis is systematically limited, rather than statistically limited. Using only one detector for this analysis also avoids introducing systematic uncertainties from combining multiple sets of data from different detectors.

\begin{figure}[t] 
\includegraphics[width=3.25in]{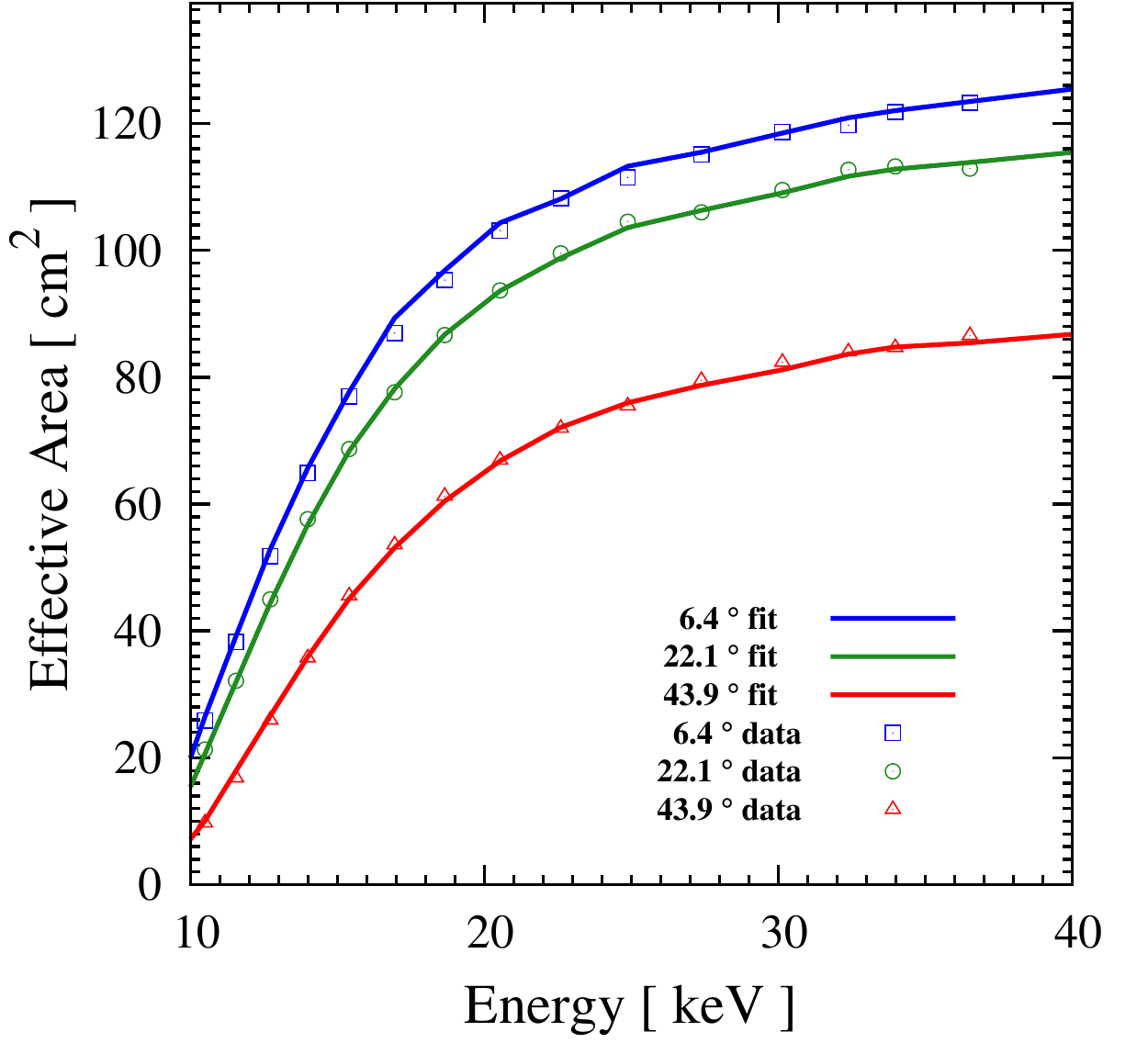} 
\caption{The effective area versus the incident photon energy for three incident angles.  The points are obtained from GBM calibration files and the lines are linear interpolations of the points. \label{fig:effa_energy}}
\end{figure}

The NaI detectors have a wide FOV, as seen in Fig.~\ref{fig:effa_angle}, which shows the effective area versus the detector zenith angle, $\theta$, for the det-7 detector.  We obtain the GBM effective area data from detector response matrix files~(\texttt{GS-008}).  Each file contains the effective area as a function of energy, for a specific detector zenith and azimuthal angle.  In Fig.~\ref{fig:effa_angle}, each point denotes the effective area extracted versus the corresponding zenith angle from the detector response file.  Beyond about $40^{\circ}$, we observe anomalous dips in certain azimuthal direction at \emph{all} energies, which is presumably caused by blockages from satellite components in the FOV of the detector.  We remove these anomalous dips by requiring adjacent bins deviate no more than $\sim 10\%$.  After this procedure, the angular dependence of the effective area can be well-described by cosine functions, as shown by the solid lines.

In Fig.~\ref{fig:effa_energy}, we show the energy dependence of the effective area for three representative zenith angles.  The points are obtained from the detector response matrix files, and are chosen from a specific azimuthal angle at which the detector FOV is not blocked.  Using the cosine fits described above, the energy dependence is obtained by linear interpolation of the model in energy.  Thus, we obtain an azimuthally symmetric model of the effective area~(i.e., it depends only on zenith angle and energy).  Shown in solid lines in Fig.~\ref{fig:effa_energy}, are the model for the given zenith angles.  The FOV blockages slightly reduce the detector sensitivity towards a particular azimuthal angle.  This effect, however, is not expected to introduce spurious spectral features in energy spectrum, since the blockages affect all energies.  We thus neglect these blockages and use the smooth angular fits to model the expected signal in the next section.

One important feature of the NaI detectors is that they are limited in their photon-tracking capabilities, i.e., one cannot simply obtain the photon flux as a function of the incidence direction for a specific position on the sky.  In other words, the tradeoff for the large FOV is poor angular resolution. Earth occultation techniques can be employed to obtain photon direction for point source studies~\cite{2012ApJS..201...33W, 2014A&amp;A...562A...7R}, but this technique has not yet been demonstrated for diffuse emissions.  Fortunately, the lack of photon tracking is not very problematic for sterile neutrino dark matter decay searches due to the large angular extent of the expected emission.  However, this does mean that one cannot accurately construct an intensity sky map of the GBM data~(Eq.~(\ref{eq:Intensity})).  As a result, one needs to properly model the signal taking into account the detector response to match the observable. In this case, the instrumental observable is counts rate~(number of photons per second), as a function of the NaI detector pointing direction.

\subsection{Expected Signal Modeling \label{sec:modeling}}
Given the sterile neutrino decay photon intensity ${\cal I}(\psi,E)$, we compute the expected number of photons, $\nu_{i,j}$, for energy bin $i$ and detector sky-pointing direction $j$.  The expected number of photons per observing time, $T_{j}$, from a particular detector pointing direction is then 
\begin{eqnarray}
 \frac{d\nu_{i,j}}{dT_{j}} &=&\int_{E^{\rm min}_{i}}^{E^{\rm max}_{i}} dE\int_{2\pi} d\Omega(\theta) \int d\tilde{E}\, \\  
\nonumber && \Bigg\{ {\cal I}(\psi,\tilde{E}) \, G\left(E,\tilde{E}\right) \, A_{\rm eff}(\tilde{E}, \theta)   \Bigg\} \, , 
\end{eqnarray}
where $E^{\rm max}_{i}$ and $E^{\rm min}_{i}$ are the boundaries of the energy bin~$i$.  We integrate over the hemisphere the NaI detector points at, i.e., over the detector zenith angle $\theta$, and attribute all the photons to pixel $j$.  A position on the sky with an angle relative to the GC, $\psi$, is related to the detector zenith angle and the pixel that the detector points at through $\psi \rightarrow \psi(\theta, j)$\,.  The pointing direction of the detector is therefore defined by $\psi(0,j )$.  The factor $G(E,\tilde{E})$ takes into account the energy resolution of the NaI detector, which we model as a Gaussian with width given by the pre-launch calibrations~\cite{Bissaldi:2008df, Meegan:2009qu}. The energy resolution is about 10\% for our analysis range.  And lastly,  $A_{\rm eff}(E, \theta)$ is our NaI detector effective area model, which is a function of energy and the detector zenith angle, as in Figs.~\ref{fig:effa_angle}, \ref{fig:effa_energy}.

Using the Dirac-delta function for the energy spectrum, the expected signal is
\begin{eqnarray} \label{eq:signal_model}
\frac{d\nu_{i,j}}{dT_{j}}(m_{s}) &=&   \frac{\rho_\odot R_\odot} {4\pi m_s \tau} \int_{E^{\rm min}_{i}}^{E^{\rm max}_{i}}dE \Bigg\{ \\
 && {\cal N}\left(\frac{m_{s}}{2}\right) \tilde{{\cal J}}\left(\frac{m_{s}}{2},j\right) G\left(E,\frac{m_{s}}{2} \right) + \nonumber \\
 && R_{\rm EG} \int d\tilde{E}  \int d\Omega\,A_{\rm eff}\, G(E,\tilde{E}) \frac{{\rm \Theta}(\frac{m_{s}}{2} -\tilde{E})}{\tilde{E} h( \tilde{E} )} \Bigg\} \, , \nonumber
\end{eqnarray}
where $\tilde{{\cal J}}\left(E,j\right)$ is the ``convolved J-factor", ${\cal N}(E)$ is a normalizing factor, $h(E) =  \sqrt{\Omega_{\Lambda} + \Omega_{M}(m_{s}/(2E))^{3}}$, $\Theta$ is the Heaviside step function.  We have suppressed the argument of $A_{\rm eff}$ for simplicity. 

\begin{figure}[t] 
\includegraphics[width=3.25in]{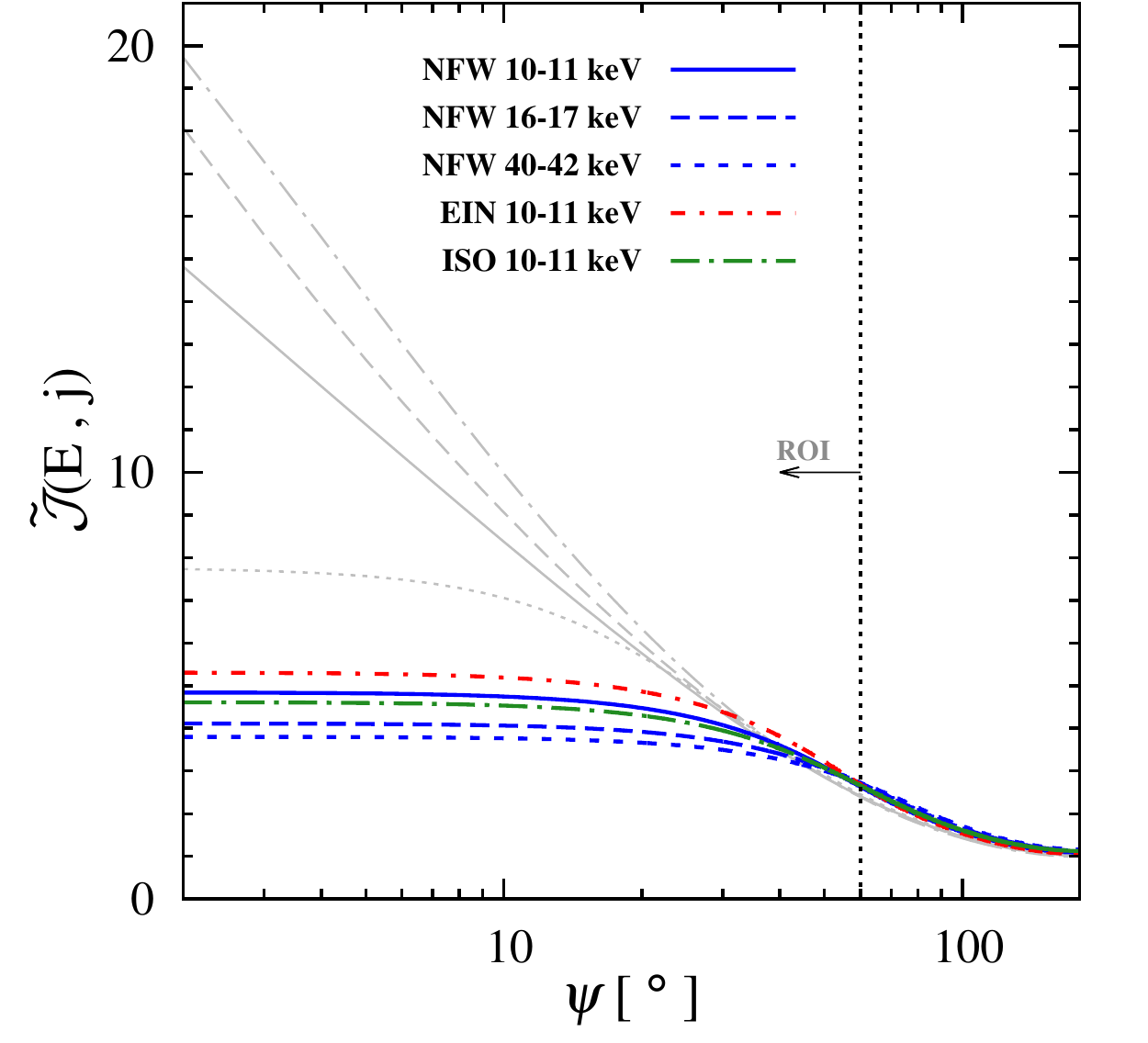} 
\caption{The convolved J-factor (Eq.~(\ref{eq:eta})) versus the opening angle with respect to the GC, $\psi(0, j)$, defined by where the detector normal is pointing.  The difference between different profiles is drastically reduced.  A small energy dependence is introduced from the effective area.  The vertical dotted line denotes the boundary of our ROI.  Shown in grey are the theoretical J-factors from Fig.~\ref{fig:Jfactor}.
} \label{fig:eta}
\end{figure}

The convolved J-factor is defined as
\begin{equation}\label{eq:eta}
\tilde{\cal{J}}(E,j) = \frac{\int {\cal J}(\psi)\, A_{\rm eff} (E,\theta ) \, d\Omega(\theta)}{ {\cal N}(E) } \, ,
\end{equation}
which takes into account the effect of detector response; it represents the J-factor defined by detector pointing directions. It depends on the detector pointing direction through $\psi(\theta, j)$.  The normalization factor ${\cal N}(E) \equiv 2\pi A_{\rm eff}(E,0)$ captures the energy behavior of the effective area. The normalization of this factor is unimportant as it cancels itself when obtaining dark matter decay fluxes/limits.  In Fig.~\ref{fig:eta}, we compare the convolved J-factor with the normal J-factor defined in Eq.~(\ref{eq:jfactor}).  Once the detector response is taken into account, the difference between different profiles decreases drastically even for pointing directions very close to the GC.  E.g., for 10--11\,keV bin, the difference in the convolved J-factor between NFW versus EIN and ISO is $\lesssim 1\%$. Therefore, systematic uncertainties due to the choice of dark matter profile are minimal.  

In the left column of Fig.~\ref{fig:counts_rate_map}, we show the modeled dark matter maps for the NaI detector from the Milky Way halo for several energies.  We pixelate the sky into 768 pixels of equal solid angle using the HEALPix scheme\footnote{http://healpix.jpl.nasa.gov~\citep{Gorski:2004by}}, i.e., each pixel corresponds to a solid angle of $ \Delta{\Omega} \simeq 1.6\times 10^{-2}$\,sr.  We use the Milky Way contribution from Eq.~(\ref{eq:signal_model}), which takes into account detector energy and angular response.  The extragalactic component only adds a constant value to the signal map.  We choose the line energy to be at the center of the chosen energy bin.  The decay rates for the sterile neutrino scenarios are chosen to approximately match the count rates of the corresponding data maps (right column, described below).  By construction, the Milky Way dark matter contribution is spherically symmetric, and the large angular extent of the signal is due to the large FOV and poor angular resolution of the NaI detectors.

\section{Data selection and reduction \label{sec:observations}}
In this section, we describe the data reduction procedures to improve the data quality and the cuts designed to reduce various backgrounds.  At the end we obtain a data set that can be compared to Eq.~(\ref{eq:signal_model}) to obtain limits for sterile neutrino dark matter decay. 

We use GBM daily data from 12-AUG-2008 to 31-DEC-2012, a total of 1601 days.  We use the \texttt{CSPEC} data\,(\texttt{GS-002}) with nominal $4.096$\,s time resolution and 128 channels in energy from 5 to 1402\,keV~(the first and last few energy bins are not usable). We then devise several cuts to improve the data quality. The goal is to obtain a data set that is representative of the diffuse sky emission as observed by the GBM NaI detectors, while minimizing various types of backgrounds.  The most dominant source of background is due to cosmic rays interacting with the satellite, directly activating the detector or triggering the detector through delayed radioactive decays of the satellite material.  

To this end, we employ the following cuts:
\begin{itemize}
\item \emph{LAT cut.}

We first select data sets that are suitable for analysis using data flags from Fermi-LAT \texttt{weekly photon} files: \texttt{LAT\_CONFIG=1}, \texttt{LAT\_MODE=5, DATA\_QUAL=1, ROCK\_ANGLE$<$50, SAA=F}. The first three conditions ensure the detector configuration and data quality are suitable for scientific analysis. The fourth condition ensures that the Earth is not in front of the LAT's FOV, which is approximately, but not exactly, the FOV of the NaI detector (we address this in the \emph{Earth cut} below). The last condition excludes the times when the satellite is inside the South Atlantic Anomaly (SAA), where the high cosmic ray activity significantly increases the radioactivity of the satellite. The GBM detectors are turned off during SAA passage, hence the observed counts are zero in these time periods.
\end{itemize}

The LAT cuts alone, however, are insufficient for reducing background events, because of the different physical locations of the detectors on the satellite, different backgrounds, and the different technologies of the LAT and the GBM. We therefore develop new cuts specifically for  the GBM.

\begin{figure*} 
\begin{center}
\includegraphics[width=7.0in]{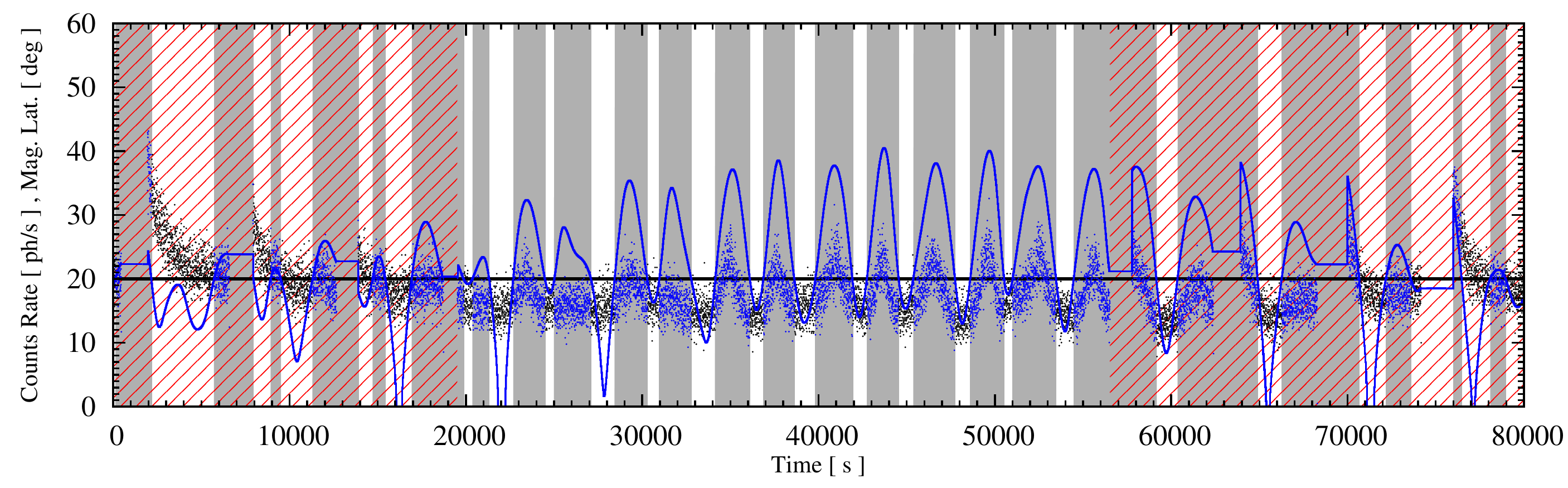} 
\caption{ A sample of the raw photon count rates for the GBM-NaI detector from 20-12-2008, from 344\,keV to 471\,keV, where the counts rate is dominated by the cosmic ray induced background.  The blue line indicates the location of the satellite in geomagnetic latitude~(in absolute value). The epochs with no data are when the satellite is physically in the SAA.  The red hashed regions represent the cuts on orbits that pass through the SAA.  The grey shaded regions illustrate the cuts on geomagnetic latitude; the data removed by this cut is colored blue.  In the end, only black data points in the white regions are used for analysis.}\label{fig:counts_angle}
\end{center}
\end{figure*}

\begin{itemize}
\item \emph{Transient sources cut.}

This cut removes the epochs when the GBM detectors detect transient sources, such as gamma-ray bursts, direct cosmic-ray hits, solar flares, Galactic X-ray transients, and magnetospheric events, etc.  Though these transients only occupy a small fraction of the observation time, some of them can be bright enough to cause the data acquisition system to overflow.  

\item \emph{Extended SAA cut.}

The \emph{LAT cut} does not completely remove events due to passages of SAA.  This is because the satellite is intensively bombarded by cosmic rays during each passage through the SAA, leaving the satellite in a highly radioactive state even \emph{after} leaving SAA. This effect is even more pronounced for consecutive passes through the SAA.  In this case, there is insufficient time for the satellite to return to its normal radioactive state. As a result, orbits passing through the SAA consecutively induce anomalously high photon count rates even when the satellite is outside the SAA.  We therefore apply cuts to remove the data collected between consecutive passages of the SAA, in addition to the times that the satellite is physically in the SAA, which are eliminated in the \emph{LAT} cut.  Removing these orbits is important to reduce events originated from cosmic rays. 

\item \emph{Earth cut.}

Lastly we apply two cuts on the orientation and the position of the NaI detector relative to the Earth.  We first require that the angle between the NaI detector normal and the vector directed from the Earth center to the satellite to be less than $50^{\circ}$. This is to reduce contamination from the Earth limb and occultation from the Earth itself.  The next cut is on the geomagnetic coordinate. The high-altitude cosmic ray activity is directly correlated to the Earth's magnetic field structure. The number of observed background events increases with geomagnetic latitude. To minimize this contamination, we select data only when the geomagnetic latitude is less than $|20|^{\circ}$.

\end{itemize}
In Fig.~\ref{fig:counts_angle}, we show an example of the data and the cuts we adopt to improve the data quality.  The data points are the counts rates on 20th December 2008 observed by det-7.  Each dot corresponds to count rates measured over $\sim 4\,{\rm s}$.  We select the energy range from 344 keV to 471\,keV, where the data is dominated by the cosmic-ray-induced background.

\begin{figure*} 
\begin{center}
\includegraphics[width=3.5in]{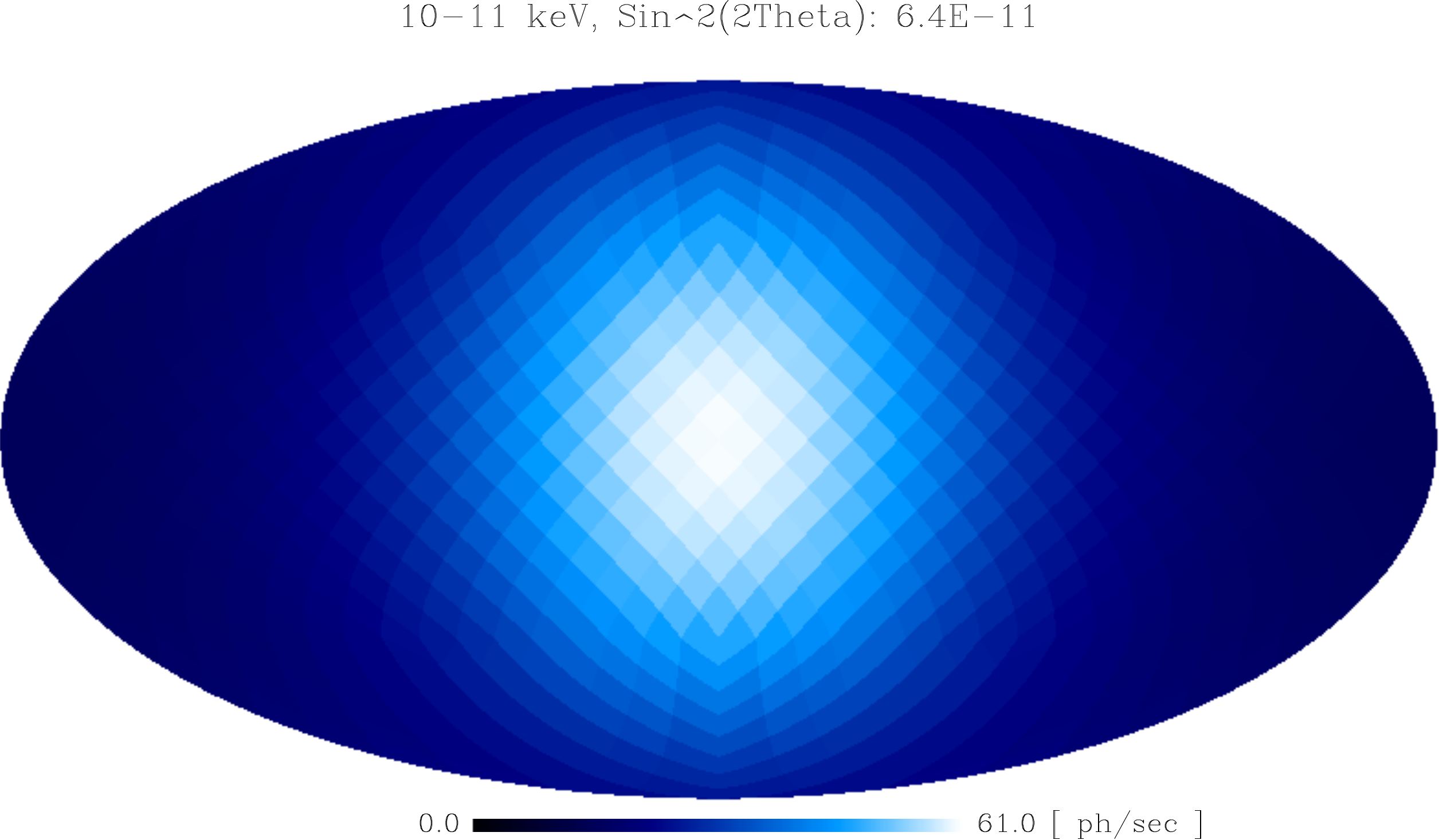} 
\includegraphics[width=3.5in]{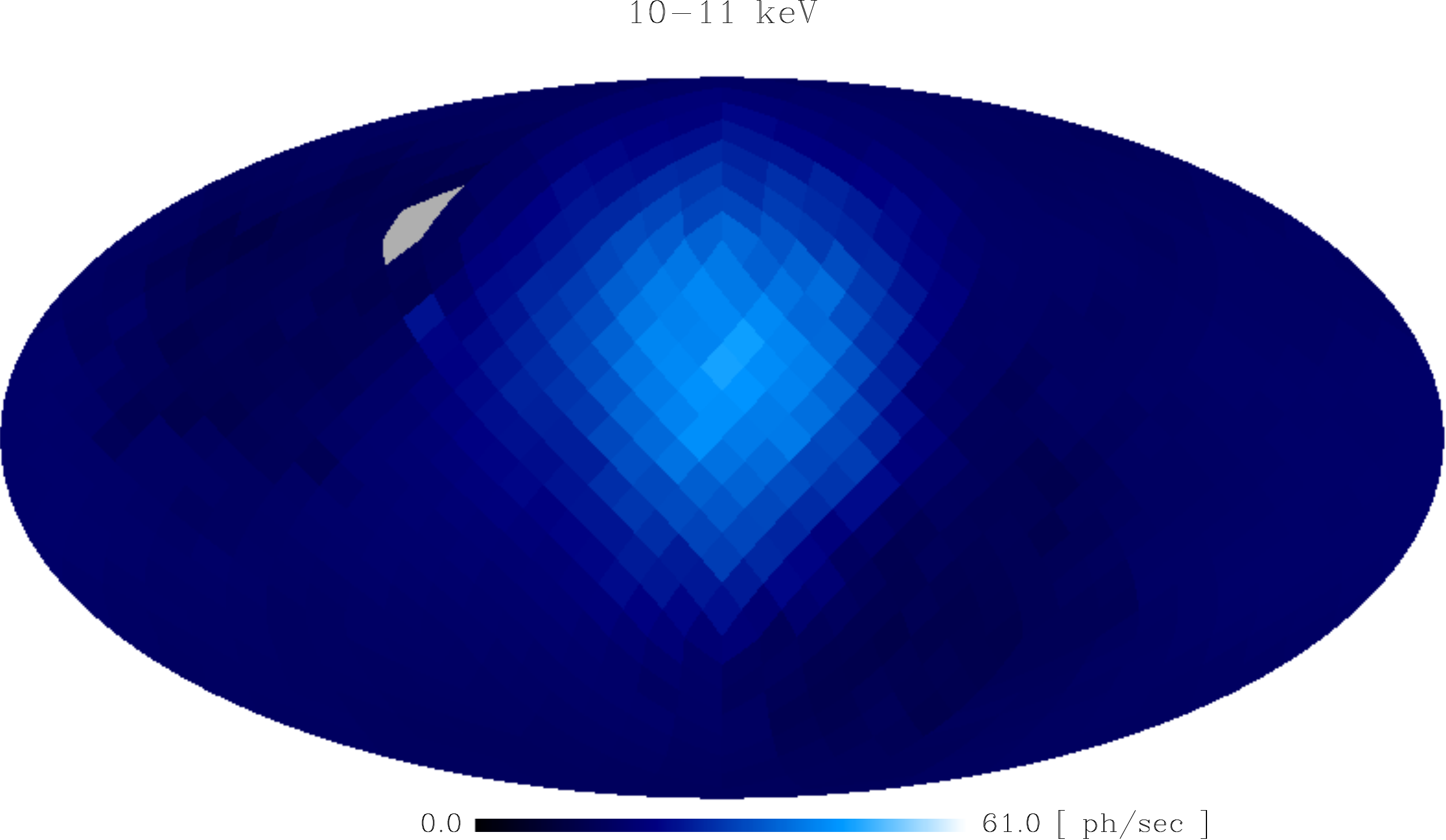} 
\includegraphics[width=3.5in]{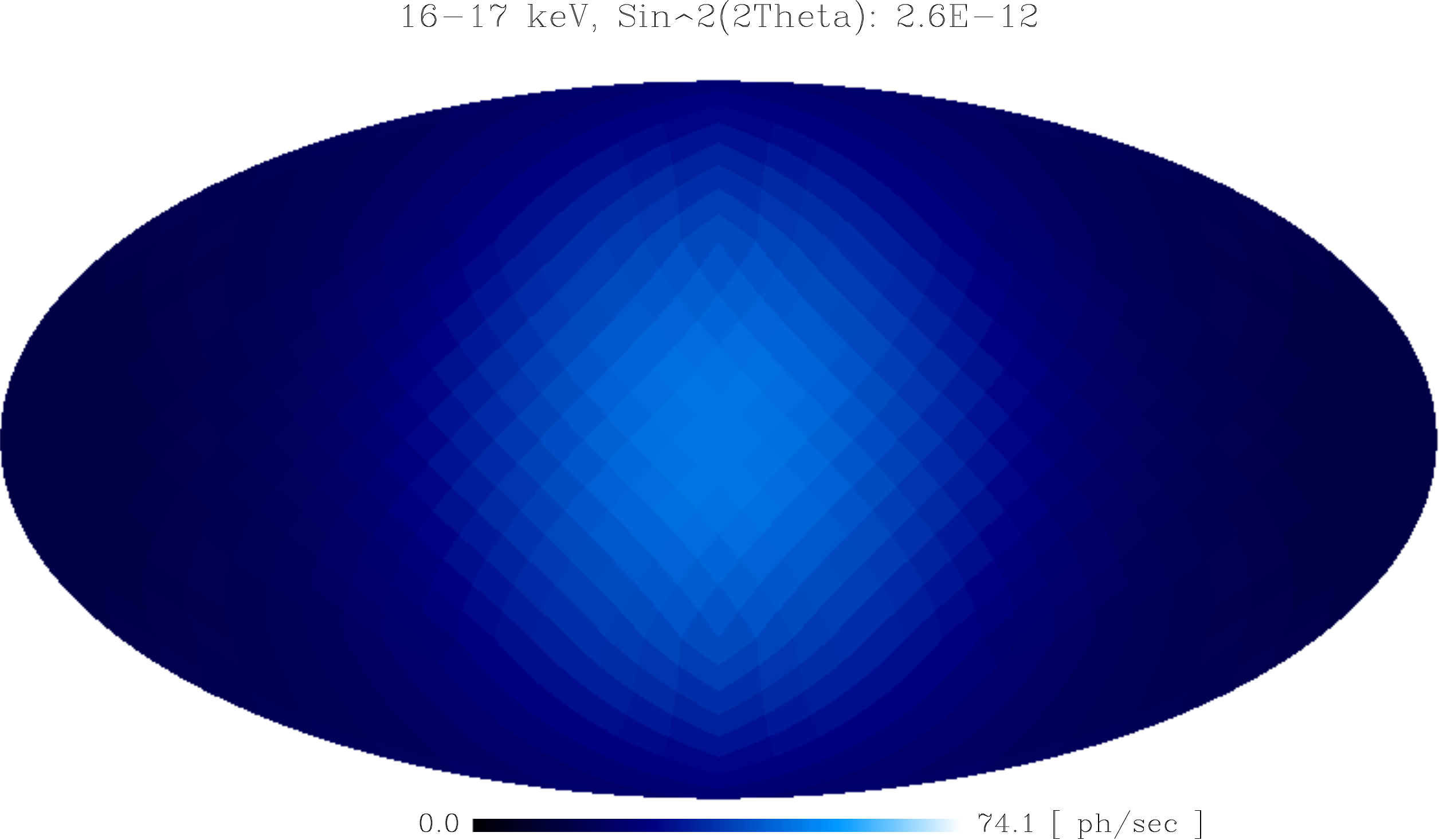} 
\includegraphics[width=3.5in]{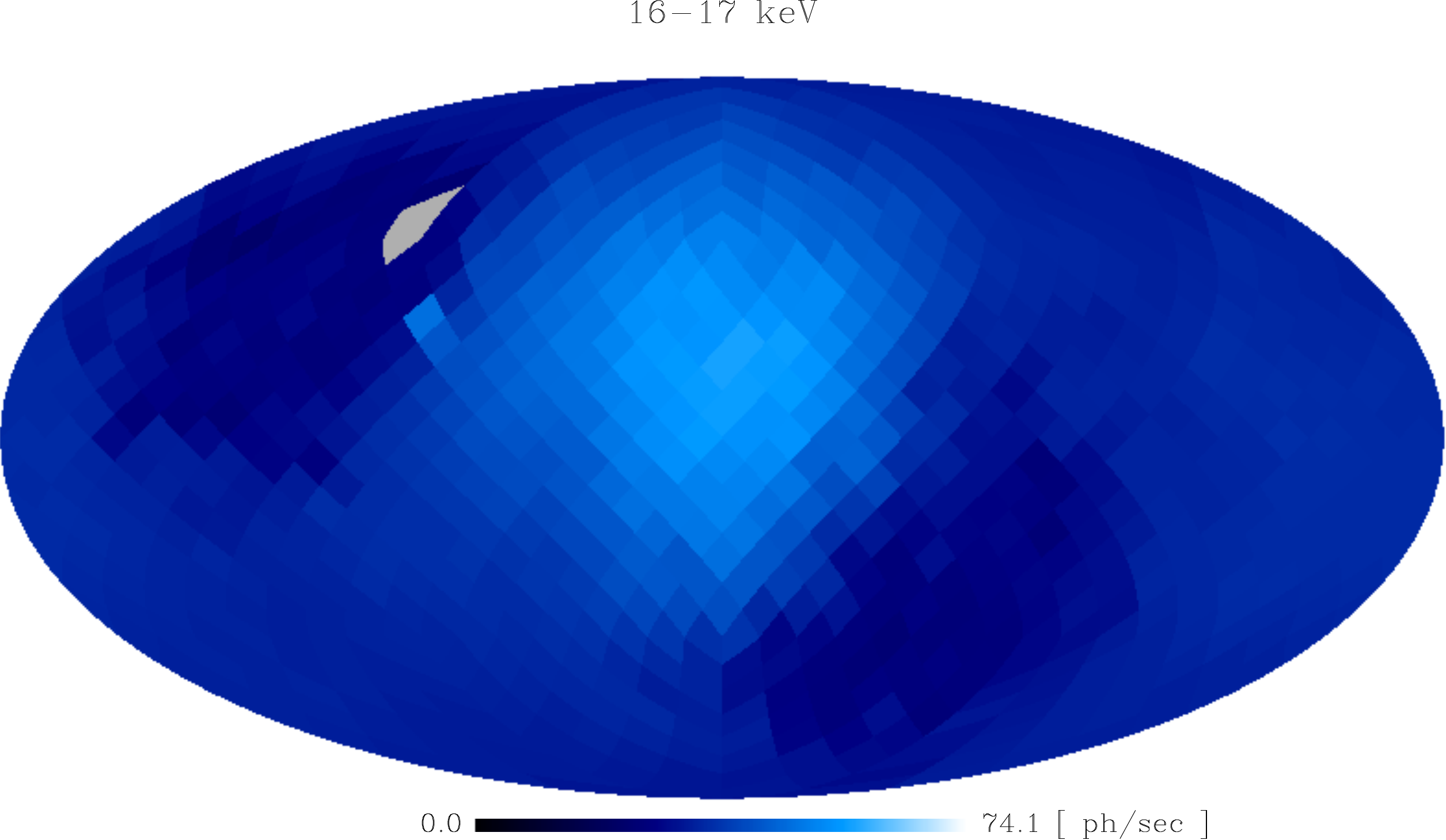} 
\includegraphics[width=3.5in]{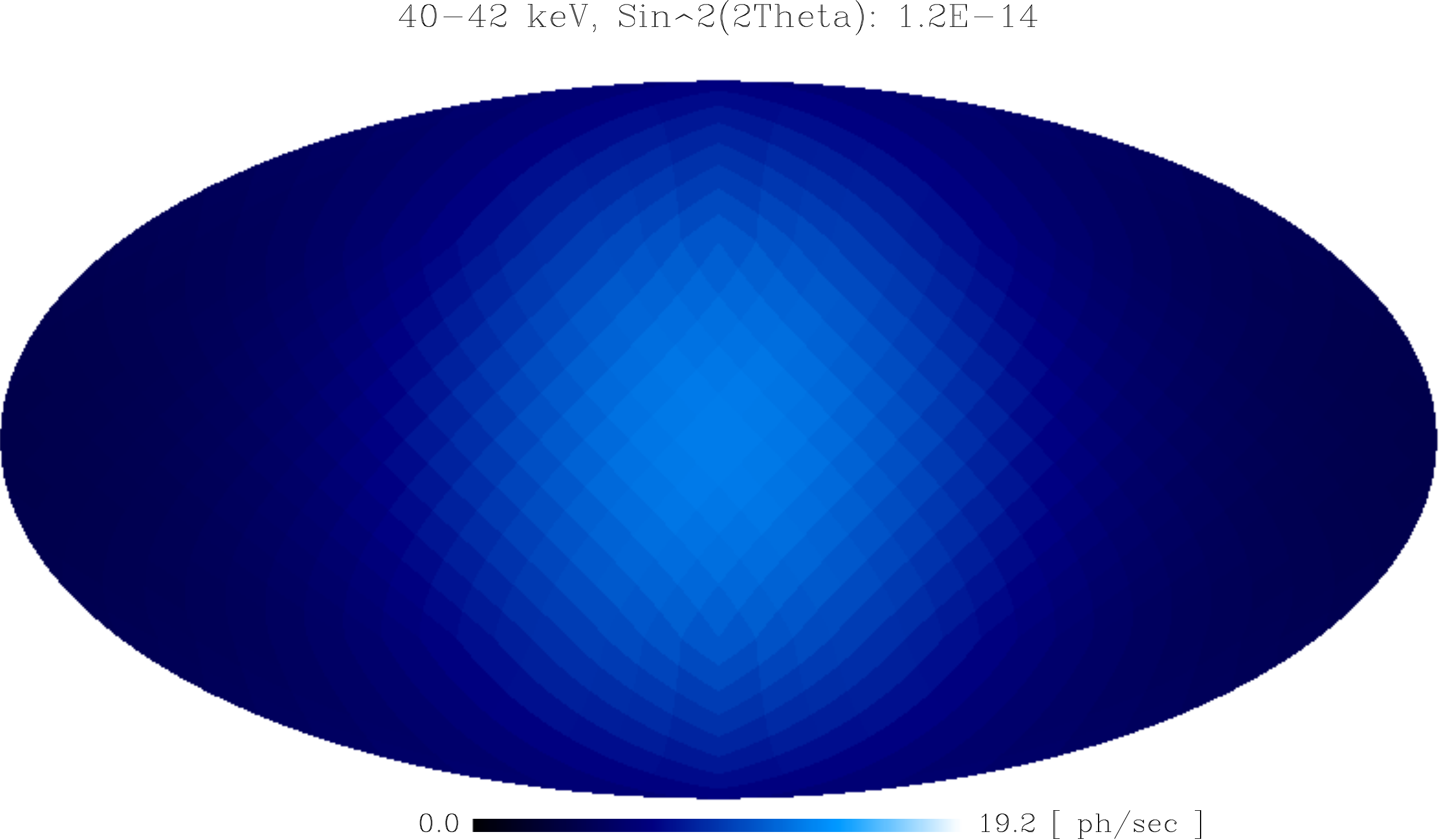} 
\includegraphics[width=3.5in]{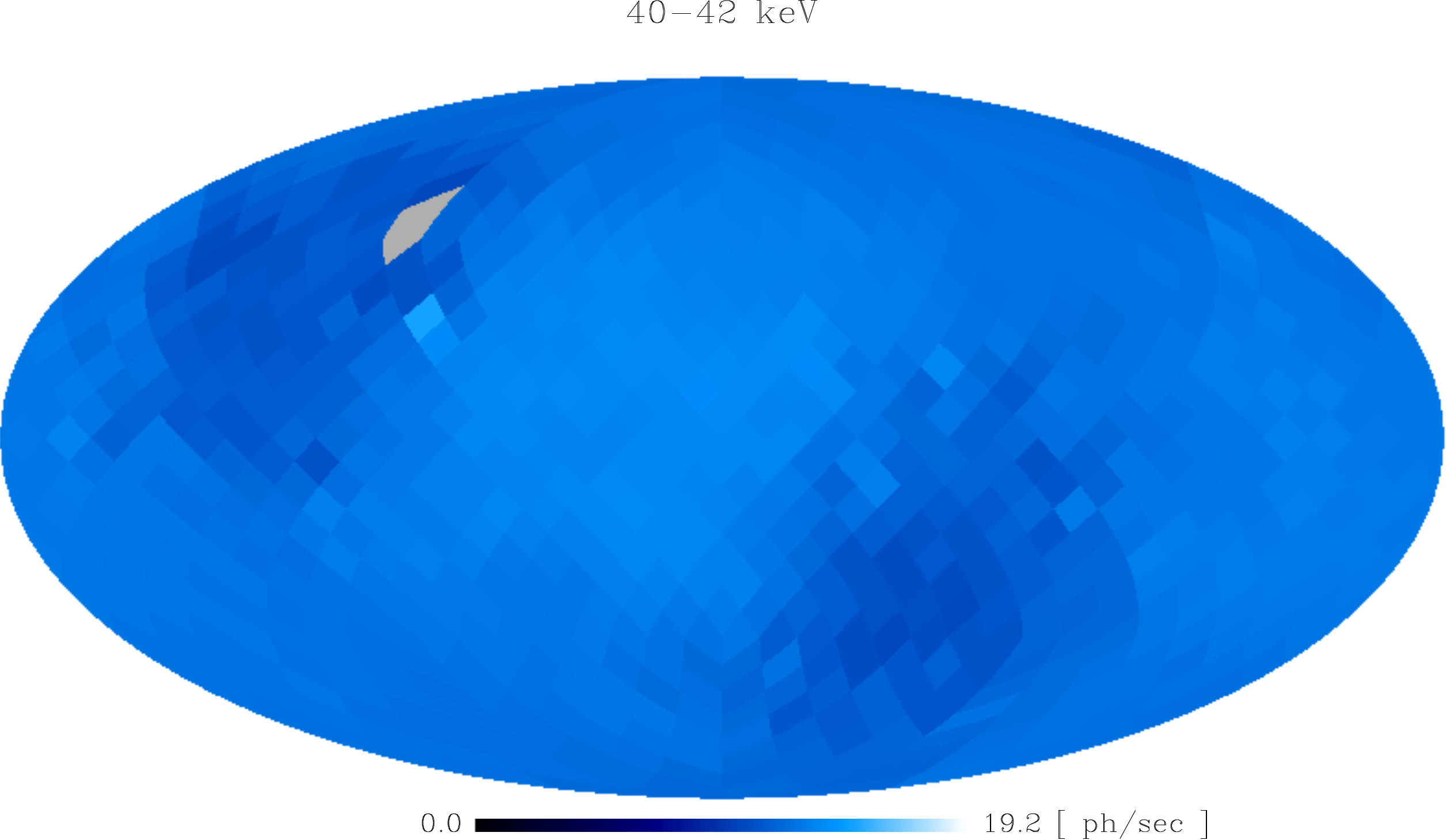} 
\caption{ ({\bf Left}) Simulated dark matter counts rate maps in Galactic coordinates for several line energies, taking into account detector response.  For each map, the assumed line energy is contained in the energy bin shown and $\sin^{2}2\theta$~(as labeled) are chosen to approximately match the observed counts in the same bin. 
({\bf Right}) The final counts rate sample from 4 years of data from the NaI detector, which corresponds to $4.6$ million seconds~($\sim 53$ days) of live time after data cuts.  All of the sky maps are pixelated into $768$ HEALPix pixels. The pixel position corresponds to the pointing direction of the detector normal.  The grey pixels are where no observing time is registered after the selection cuts.   }\label{fig:counts_rate_map}
\end{center}
\end{figure*}

The first feature that can be seen in Fig.~\ref{fig:counts_angle} is the series of epochs with no count rate.  This is because the detector was shut down when the satellite is in the SAA.  These epochs are removed in the \emph{LAT cut}.  It is also clear that the count rates are anomalously high even after the satellite leaves the SAA (i.e., right after the gap), due to the increased radioactivity of the satellite.  These epochs are removed in the \emph{Extended SAA cut}.  These two SAA related cuts are represented by the red hashed regions. 

The second feature is the the oscillatory shape during the middle of the day.  Overlaying the data points we also plot the location of the satellite in geomagnetic latitude~(blue line).  One can see the count rates are correlated with the geomagnetic latitude.  We therefore remove all the data recorded when the geomagnetic latitude is larger than $|20|^{\circ}$.  This cut is represented by the grey shaded region, and the removed data points are labelled in blue.  The choice of a uniform $|20|^{\circ}$ geomagnetic latitude cut is a balancing act between maximizing sky coverage and reducing background.  More sophisticated cuts may be possible.   

 The importance of our cuts for improving the data quality can be estimated in Fig.~\ref{fig:counts_angle}.  The increased count rate right after SAA can be a factor of a few higher.  Even the variation due to geomagnetic latitude can be up to a factor of two.  Our \emph{Extended SAA cut} and \emph{Earth cut} is therefore necessary to reveal the astrophysical component, which is comparable to the detector background at low energies~(shown below).  Transient sources does not contribute significantly to the total counts, but they can dominate a particular sky pixel.  Since they only contribute a small fraction of the live time, the \emph{Transient sources cut} is very efficient. 

The final data products obtained are observed counts and exposure time over 128 energy bins and 768 sky pixels.  The total live time of the data product is $\sim 4.6\times 10^{6}$ seconds ($\sim 53$ days). Despite having a huge reduction from the raw data, we are still far from statistically limited, as will be shown below.  

In the right panel of Fig.~\ref{fig:counts_rate_map}, we show the counts rate sky map for the labeled energy bins.  At low energies, we observe a clear excess towards the GC.  We interpret the excess as astrophysical~(i.e., non-instrumental-related) emissions from the Milky Way.  The astrophysical flux is about $\sim 10^{-1}\,{\rm cm^{-2} s^{-1}}$ if one extrapolate from the high energy observations~\cite{2005ApJ...635.1103B, 2008ApJ...679.1315B, 2011ApJ...739...29B}, which matches the observed counts rate of about $\sim 10\,{\rm s^{-1}}$.  The observed excess towards the GC also shows a small north-south asymmetry, which probably reflects the underlying distribution of diffuse and discrete X-ray sources.

For the maps at high energies, the morphology is significantly more isotropic than at low energies, with small variations that trace orbital structure, as expected from cosmic-ray-induced backgrounds.  For example, the two dark spots near the orbital pole in high energies is also seen in the low energy map.

As a result, we conclude the low energy data set consists of a mixture of astrophysical diffuse and point source emissions, plus residual cosmic-ray-induced background.  The grey pixels in the maps represent positions on the sky that were not visited by the detector, and are excluded from the analysis.

Using the data sky map and the convolved J-factor $\tilde{\cal{J}}(E,j)$, we can determine the region of interest~(ROI) for our analysis. As $\tilde{\cal{J}}(E,j)$ flattens out at small angles due to the poor detector angular resolution, as shown in Fig.~\ref{fig:eta}, there is little benefit in choosing a small ROI.  We carry out a signal-to-noise study to look for an optimal ROI angle.  The morphology of the GC excess seen in low energies turns out to be comparable to the smoothed dark matter distribution, and the signal-to-noise is fairly insensitive to the choice of angle.  This is a direct consequence of the poor angular resolution of the NaI detector.  We conservatively choose a large ROI, which consists of pixels within $60^{\circ}$ from the GC, i.e., $\psi< 60^{\circ}$.  With this selection, we have enough pixels to average out potentially spurious behavior in some individual pixels, and have more than enough statistics.  Lastly, this ROI only minimally overlaps with the dark spot positions near the orbital poles.

In Fig.~\ref{fig:counts}, we show the binned counts spectrum for the data sample in the GC ROI.  As a comparison, we also show the spectrum for the anti-GC ROI ($\psi > 120^{\circ}$).  The total observed time for the two samples are 975066\,s and 911451\,s, respectively, and this difference is the main reason the normalization differ in high energies.  In general, the counts spectrum has a power-law behavior at high energies, as expected from cosmic-ray induced background.  There are several prominent line features from excited energy levels of ${^{127}}$I at 57.6\,keV and 202.9\,keV, as well as the 511\,keV line from positron annihilation from the atmosphere and nearby materials~\cite{Meegan:2009qu}.  At low energies, the GC and anti-GC spectral shape starts to deviate, and the difference in normalization increases compared to high energies.  This indicates that the astrophysical component starts to appear in the GC sample.

\section{Limits on sterile neutrinos\label{sec:limits}}

We present two limits on sterile neutrino decay lines.  The first is a conservative limit based purely on flux comparison.  The second uses the fact that the signal is a photon line, while the background flux is approximately a power-law within the search energy window.

\subsection{Flux Analysis}

The most robust constraint one can place on the amplitude of a sterile neutrino dark matter decay signal is to require that the expected signal counts do not exceed the total measured counts.  For a set of dark matter masses, we compare, bin by bin, the predicted counts from sterile neutrino decay to the total counts measured. This approach therefore assumes the hypothesized signal dominates the observed spectrum without any assumptions about the detector and astrophysical background.

The expected signal counts are given by summing the count rates in all the individual sky pixels within the ROI, using Eq.~(\ref{eq:signal_model}), weighted by the actual observing time $T_j$ in each pixel, 
\begin{eqnarray} \label{eq:pre_fs}
\nu_i &=& \sum_{j}^{ROI} T_{j} \frac{d\nu_{i,j}}{dT_{j}} \, .
\end{eqnarray}
The measured photon counts data from all pixels in the ROI is 
\begin{equation} \label{eq:data}
d_i =  \sum_{j}^{ROI} N_{i,j}  \,,
\end{equation}
where $N_{i,j}$ is the number of counts in energy bin $i$ and pixel $j$ measured by the GBM detector.

We obtain the flux analysis limit on the decay rate, $\Gamma_{s}$, by requiring $\nu_{i} < d_{i} $ for all energy bins for each $m_{s}$.  The limit obtained this way is very conservative. It is unlikely that sterile neutrino decay, which has a sharp spectral shape, would dominate a narrow energy range in the count spectrum while other components conspire to vanish in that particular energy range.

\begin{figure}[t] 
\includegraphics[width=3.5in]{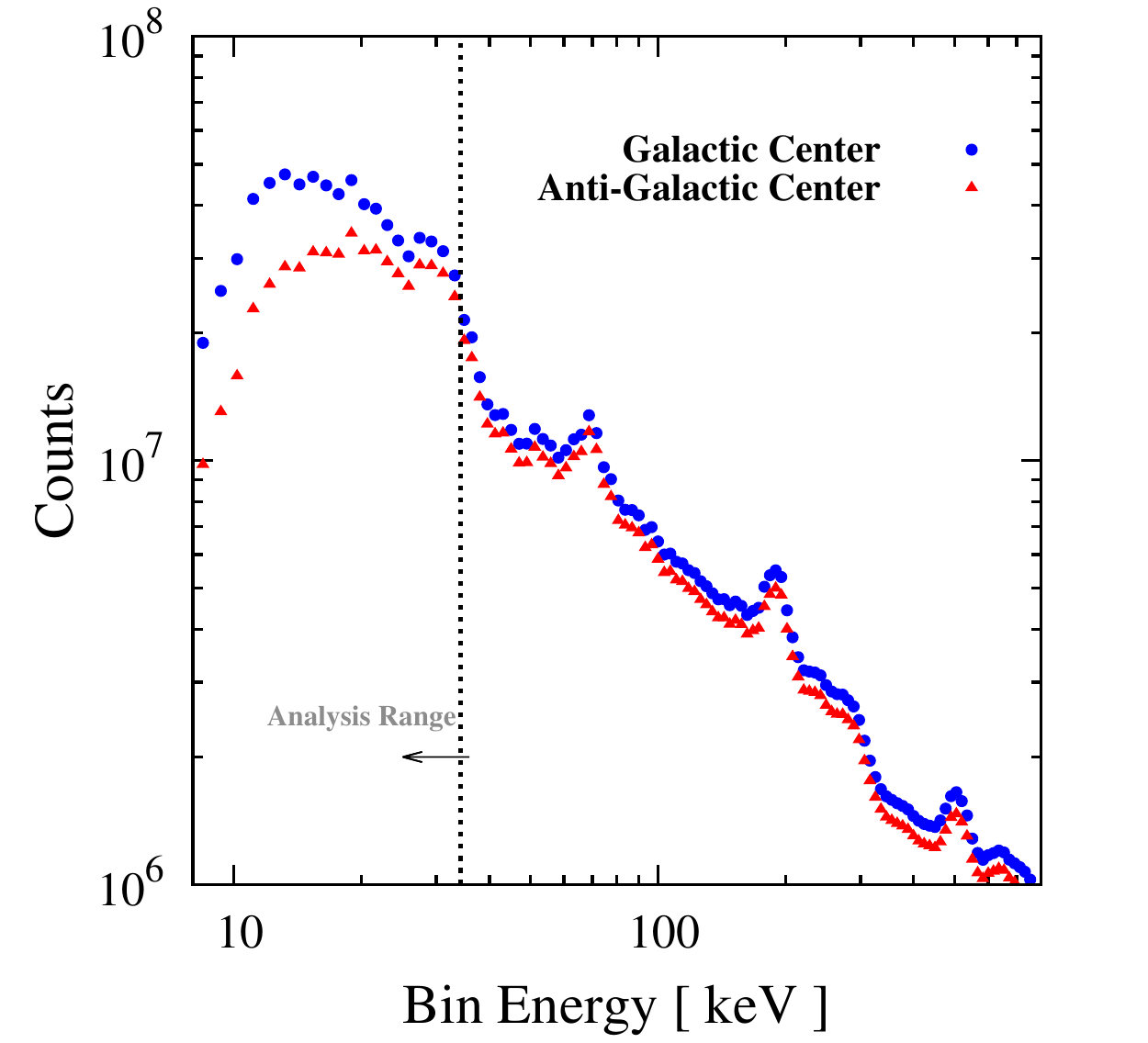} 
\caption{ The counts spectrum for the final data sample for both GC ROI\,($\psi < 60^{\circ}$) and anti-GC ROI\,($\psi > 120^{\circ}$) chosen to have the same solid angle. The dominant component is a power-law plus various background lines. The overall difference in normalization is due to the higher exposure towards the GC than the anti-GC direction. The additional excess at low energies towards the GC region suggests the rise of the astrophysical component.  The vertical dotted line indicates the energy bins used for the spectral analysis.    \label{fig:counts}}
\end{figure}

\subsection{Spectral Analysis \label{sec:spectral}}
The sensitivity to sterile neutrino dark matter decay can be improved dramatically using the observation that the sterile neutrino decay signal and the dominant background have different spectral shapes.  

\subsubsection{A simple background model \label{sec:bkhmodel}}

As shown above, our GC data sample contains an astrophysical component as well as internal detector backgrounds.  The astrophysical contribution from the inner galaxy is dominated by points sources~(all unresolved by GBM), and the energy spectrum was shown to be well described by a power law above 20\,keV~\cite{2005ApJ...635.1103B, 2008ApJ...679.1315B, 2011ApJ...739...29B}.  The internal detector background is a consequence of cosmic rays interacting with the satellite components, which retains the power-law behavior of the incoming cosmic rays.  We therefore expect the energy spectrum to have a power law distribution.

A power-law spectrum is an even better approximation when we analyze the data in small energy windows.  We consider 15 of such search windows, one for each line energy of interest.  The line energies are the corresponding energies of the energy bin number 6 to 20 in GBM numbering scheme~(labeled by $i_{0}$).  For each search bin $i_{0}$, the search window contains a number of energy bins~(labeled by $i$), where
\begin{equation}
{\rm Max} \left( i_{\rm min}, i_{0}-\Delta w  \right)< i < i_{0} +\Delta w \, . 
\end{equation}
The window size is $\Delta w = 5$, which makes the window width on each side about 3--4\,$\sigma$ of the energy resolution at the line energy.  For line energies near the low energy cutoff, we truncate the search window at the lowest usable energy bin, $i_{\rm min}= 6$, which corresponds to a central bin energy of 9.3\,keV.  The signal line energy in such case is not located in the center bin of the search window.  

With the power-law assumption for the non-dark matter components in each search window, the model photon counts spectrum therefore contains the dark matter signal component~($d\nu/dE$) and a power-law background component~($db/dE$),
\begin{eqnarray}
\frac{d\nu}{dE} &=& f_{s} \Bigg\{ \delta(E-E_{0}) {\cal N}(E) +\\
&& R_{\rm EG} \frac{\sum T_{j}}{\sum T_{j} \tilde{J}(E_{0},j)} \int A_{\rm eff}(E) d\Omega \frac{\theta(E_{0}-E)}{E h(E)} \Bigg\} \, ;  \nonumber  \label{eq:line} \\
\frac{db}{dE} &=&  \beta \left( \frac{E}{E_0} \right)^{-\gamma} {\cal N}(E) \, ,   \label{eq:bkg}
\end{eqnarray}
respectively, where $E_0$ is the energy of bin $i_{0}$.  The factor ${\cal N}(E)$ takes into account the energy response of the effective area.  The model has only three free parameters, $f_{s}, \beta$, and $\gamma$.  The factor $f_{s}$ is the amplitude of the dark matter signal, which is the only parameter that we are interested in.  The normalization and the spectral index of the background power law are thus treated as nuisance parameters, $\Xi = (\beta,\gamma)$.

\begin{figure*}[p!]
\centering
\includegraphics[angle=0.0, width=0.3\textwidth]{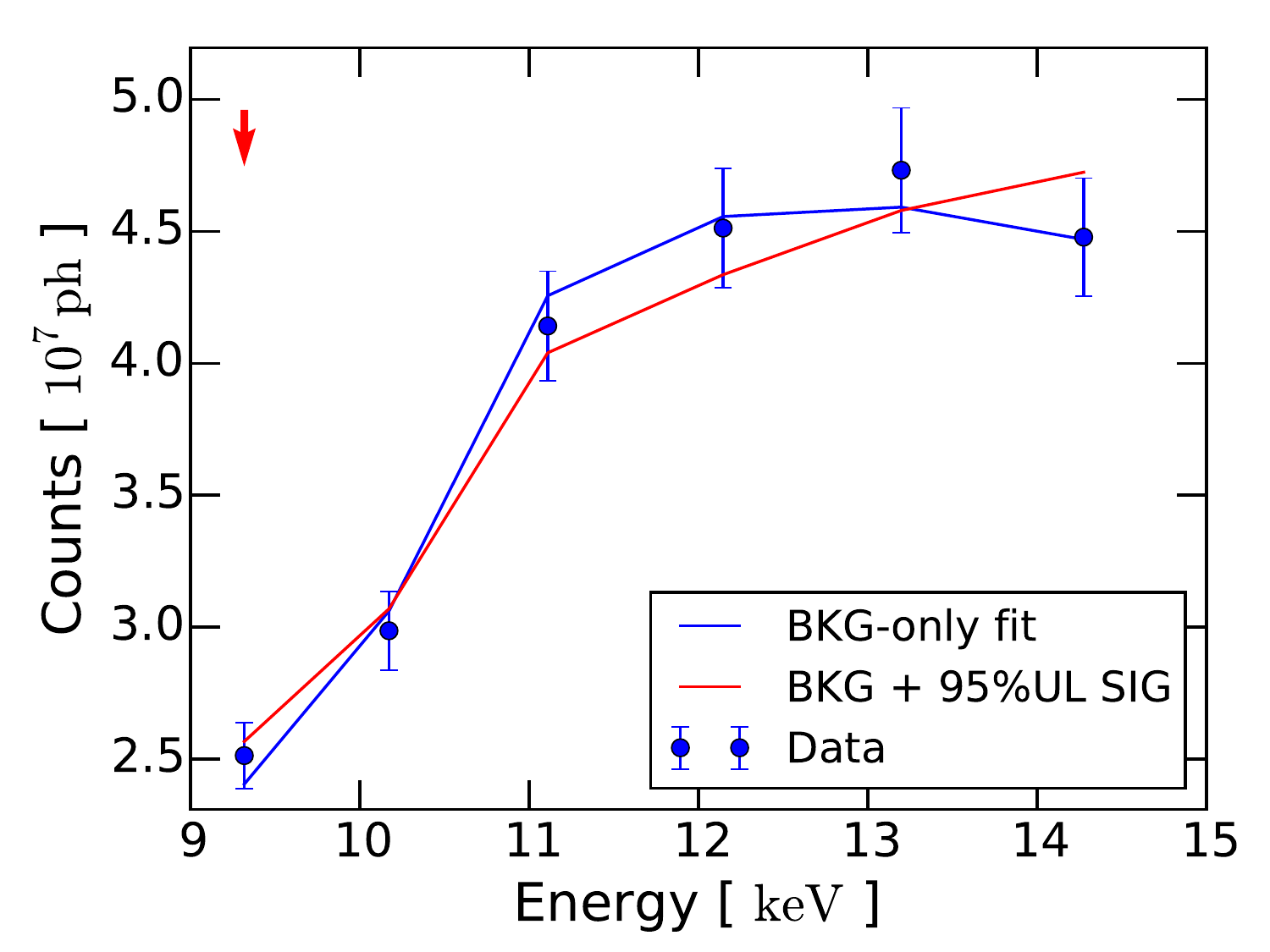}
\includegraphics[angle=0.0, width=0.3\textwidth]{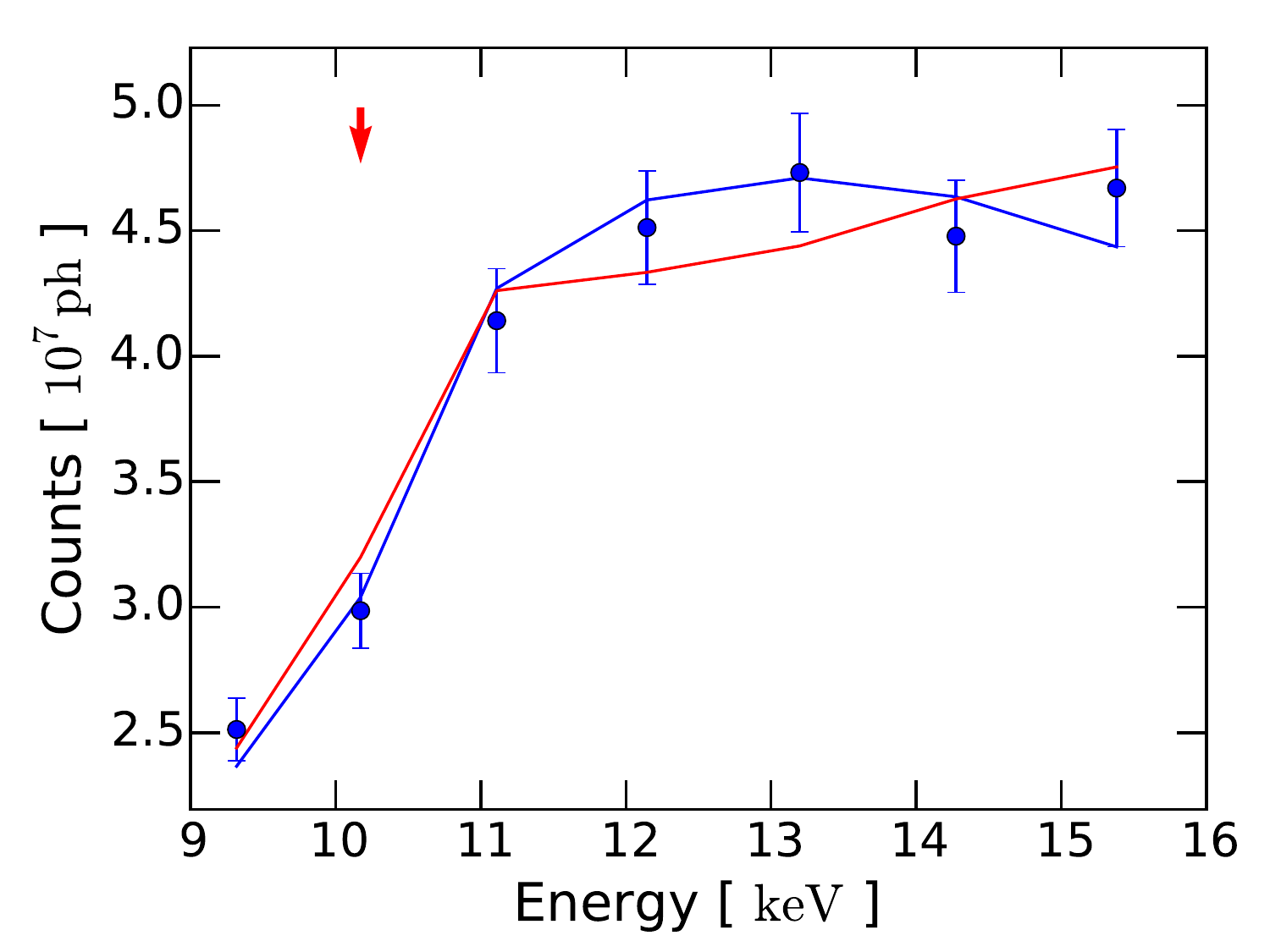}
\includegraphics[angle=0.0,width=0.3\textwidth]{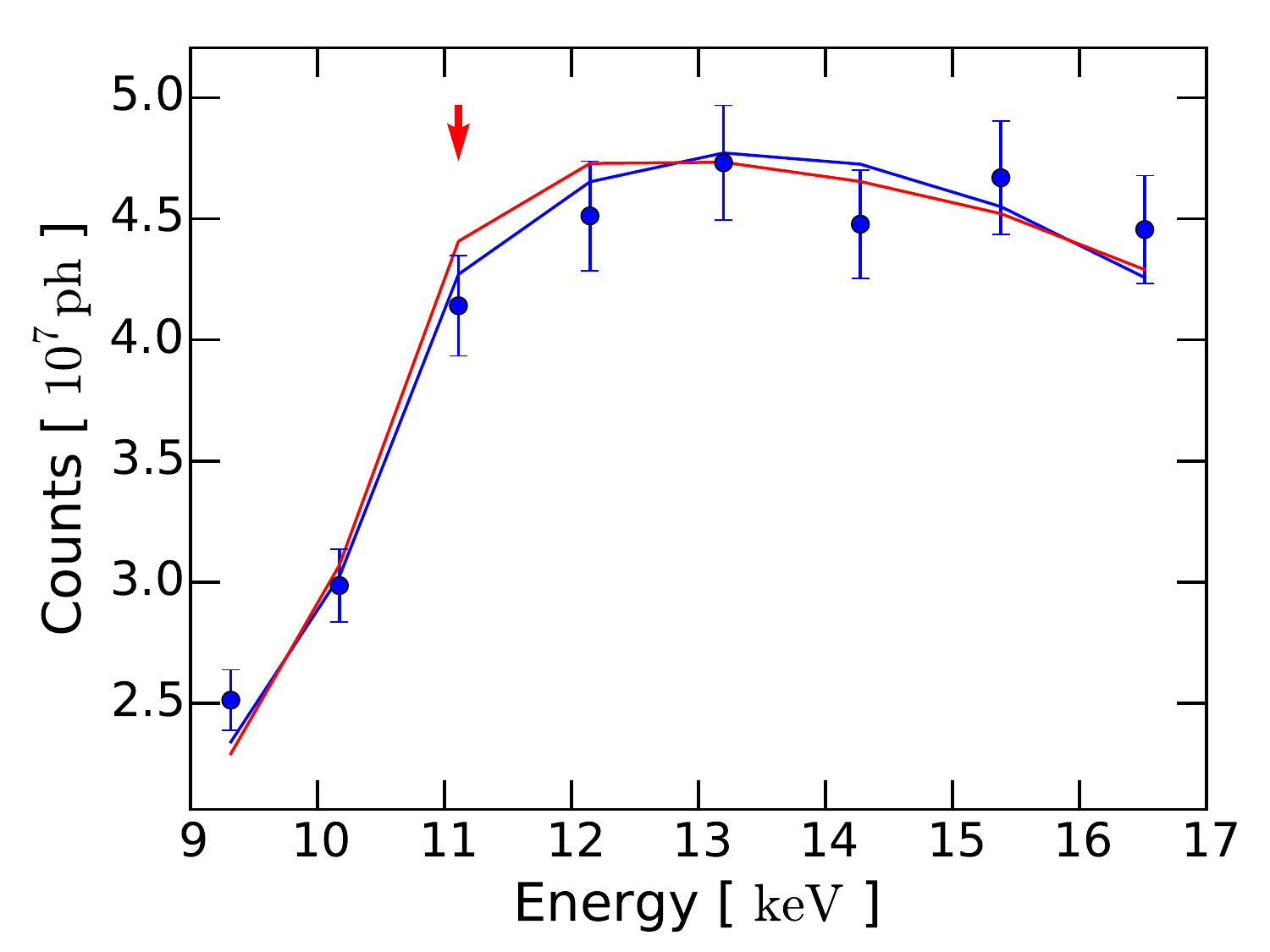}
\includegraphics[angle=0.0,width=0.3\textwidth]{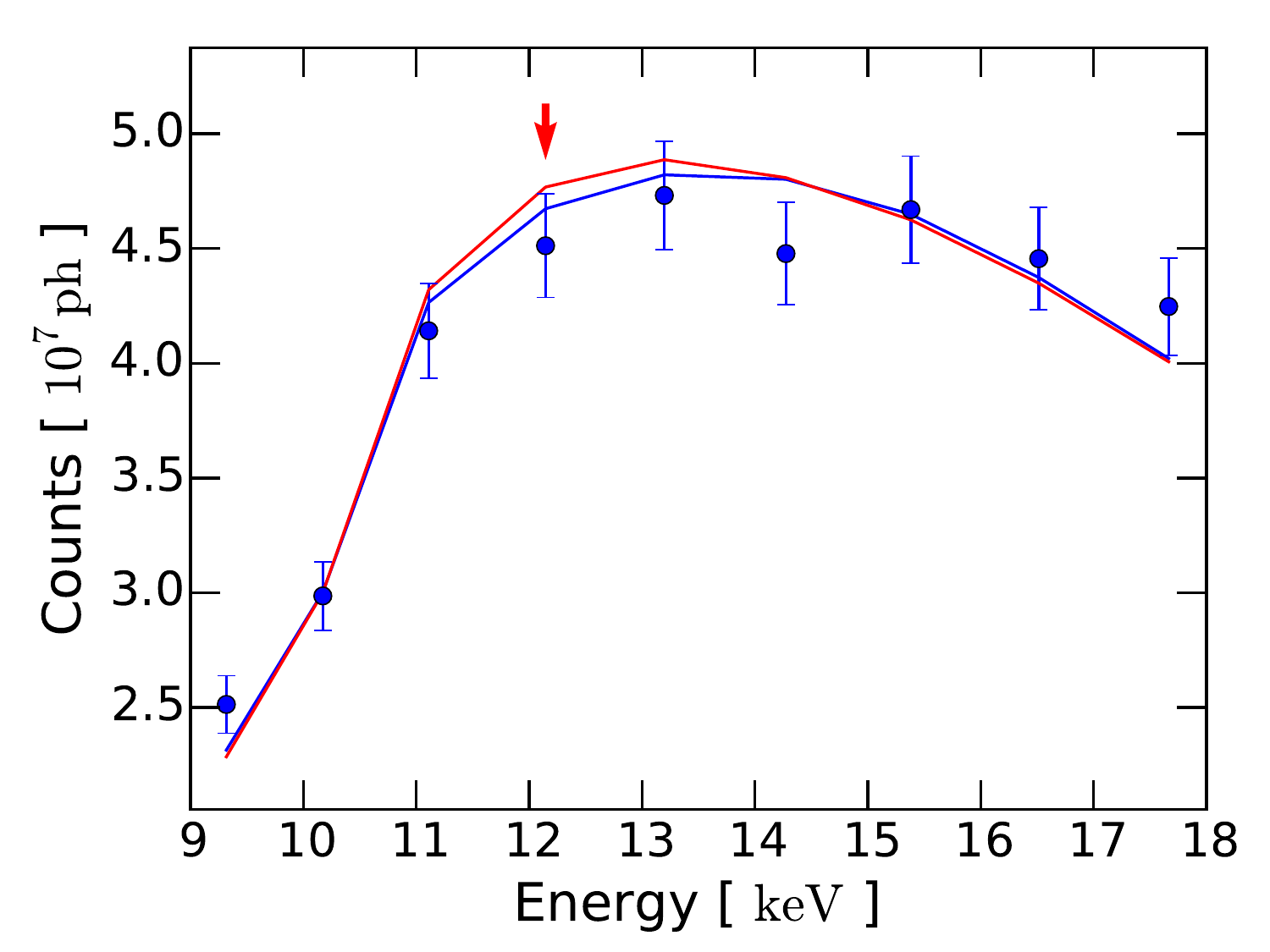}
\includegraphics[angle=0.0,width=0.3\textwidth]{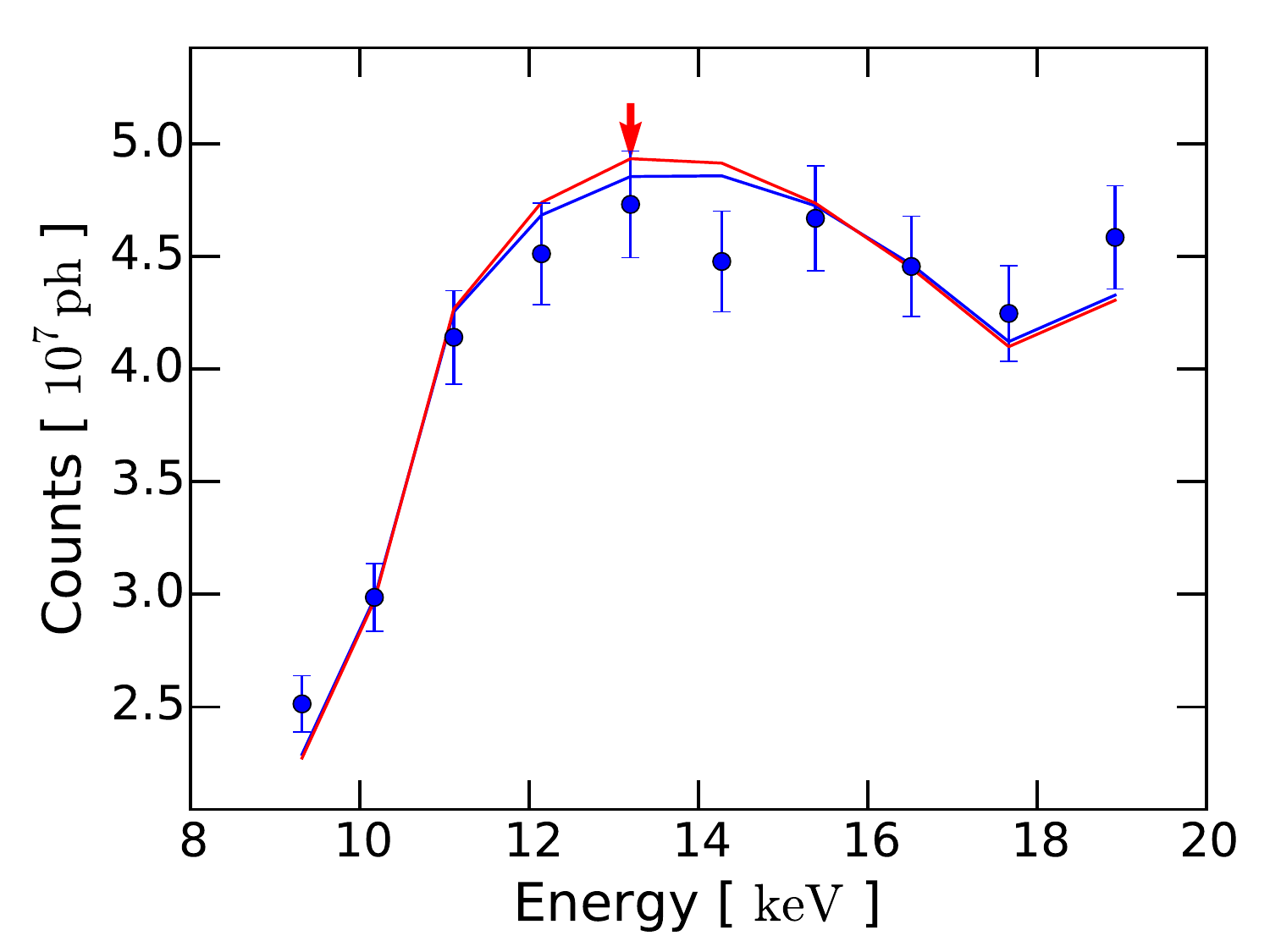}
\includegraphics[angle=0.0,width=0.3\textwidth]{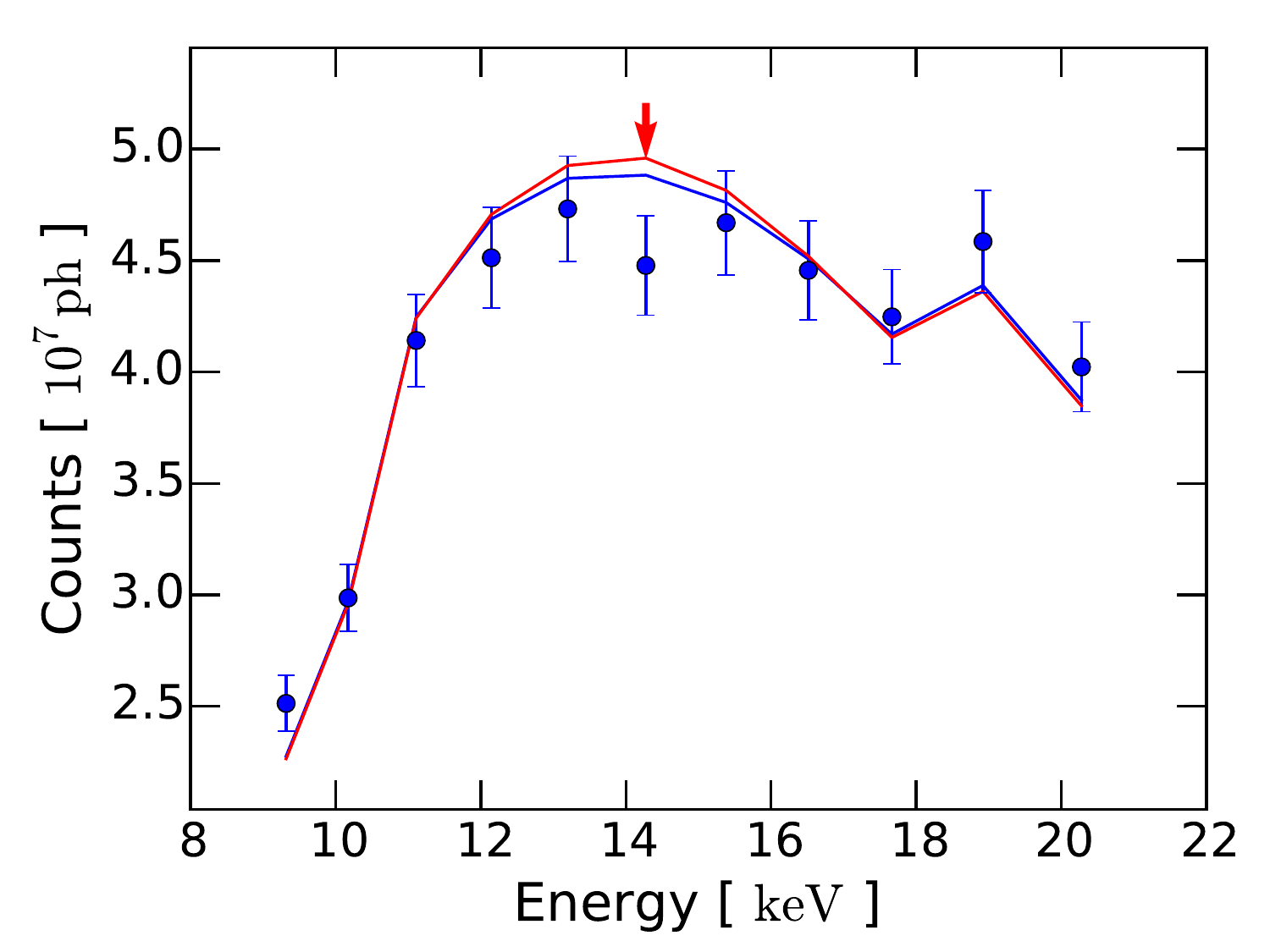}
\includegraphics[angle=0.0,width=0.3\textwidth]{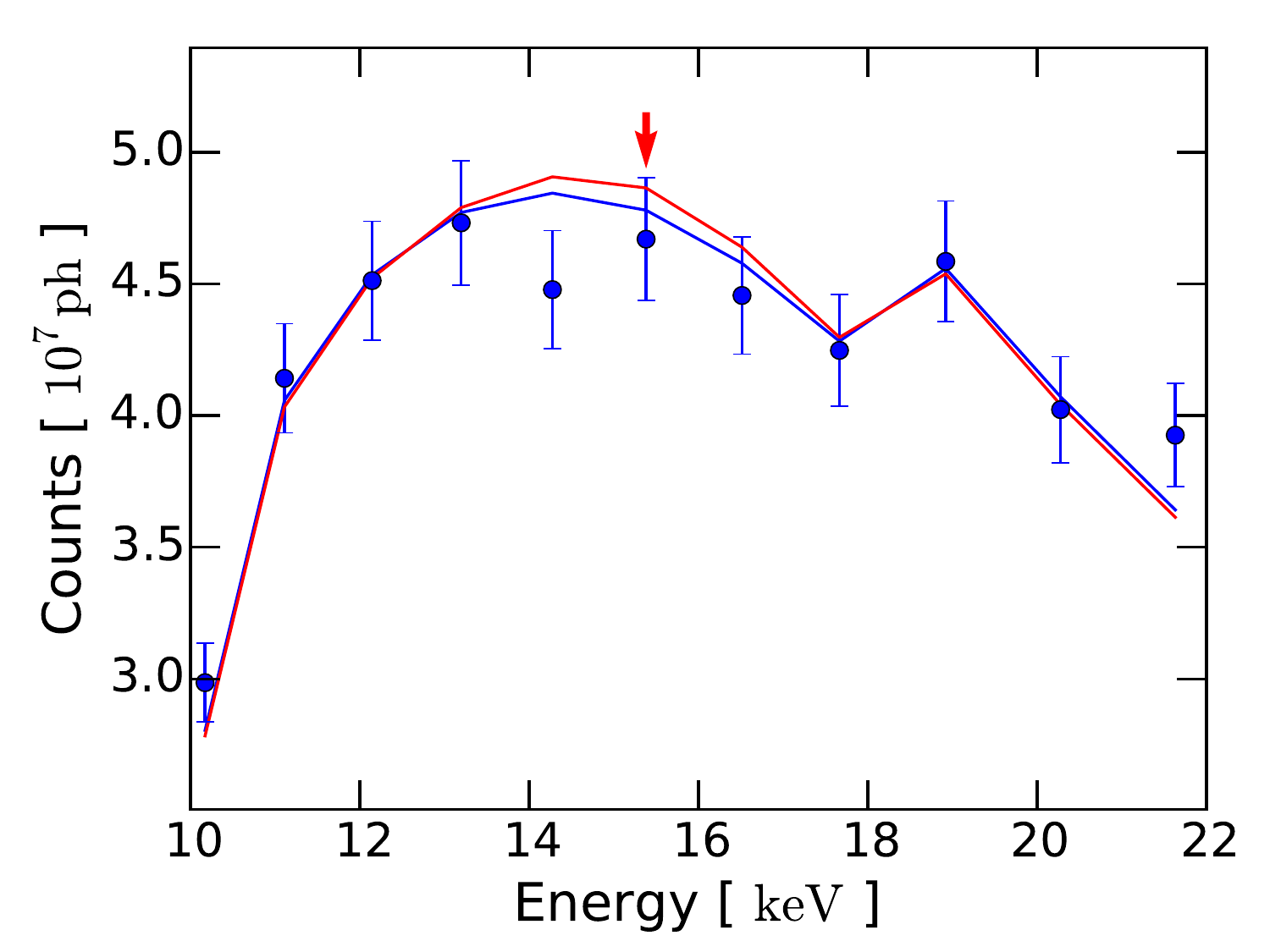}
\includegraphics[angle=0.0,width=0.3\textwidth]{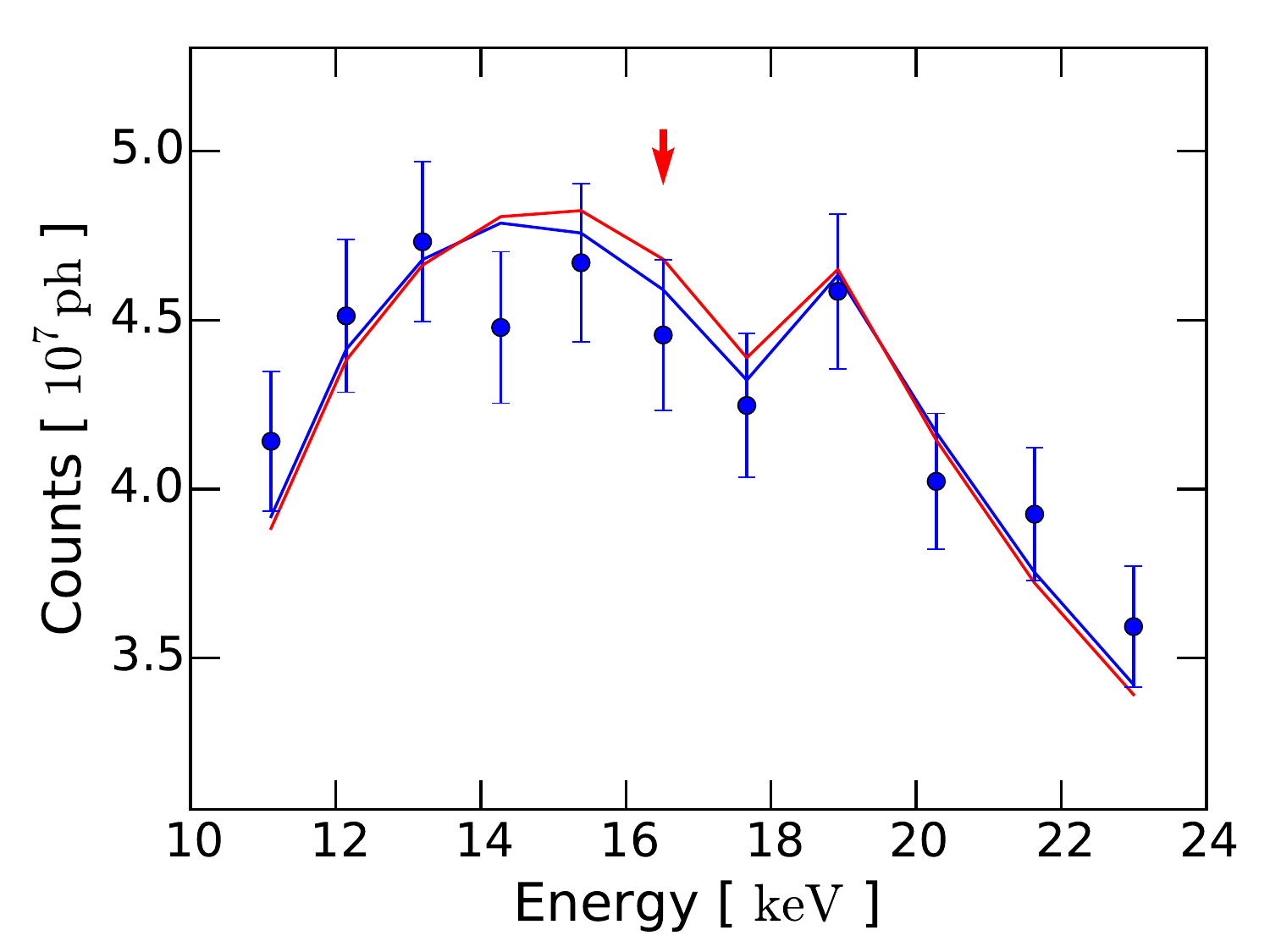}
\includegraphics[angle=0.0,width=0.3\textwidth]{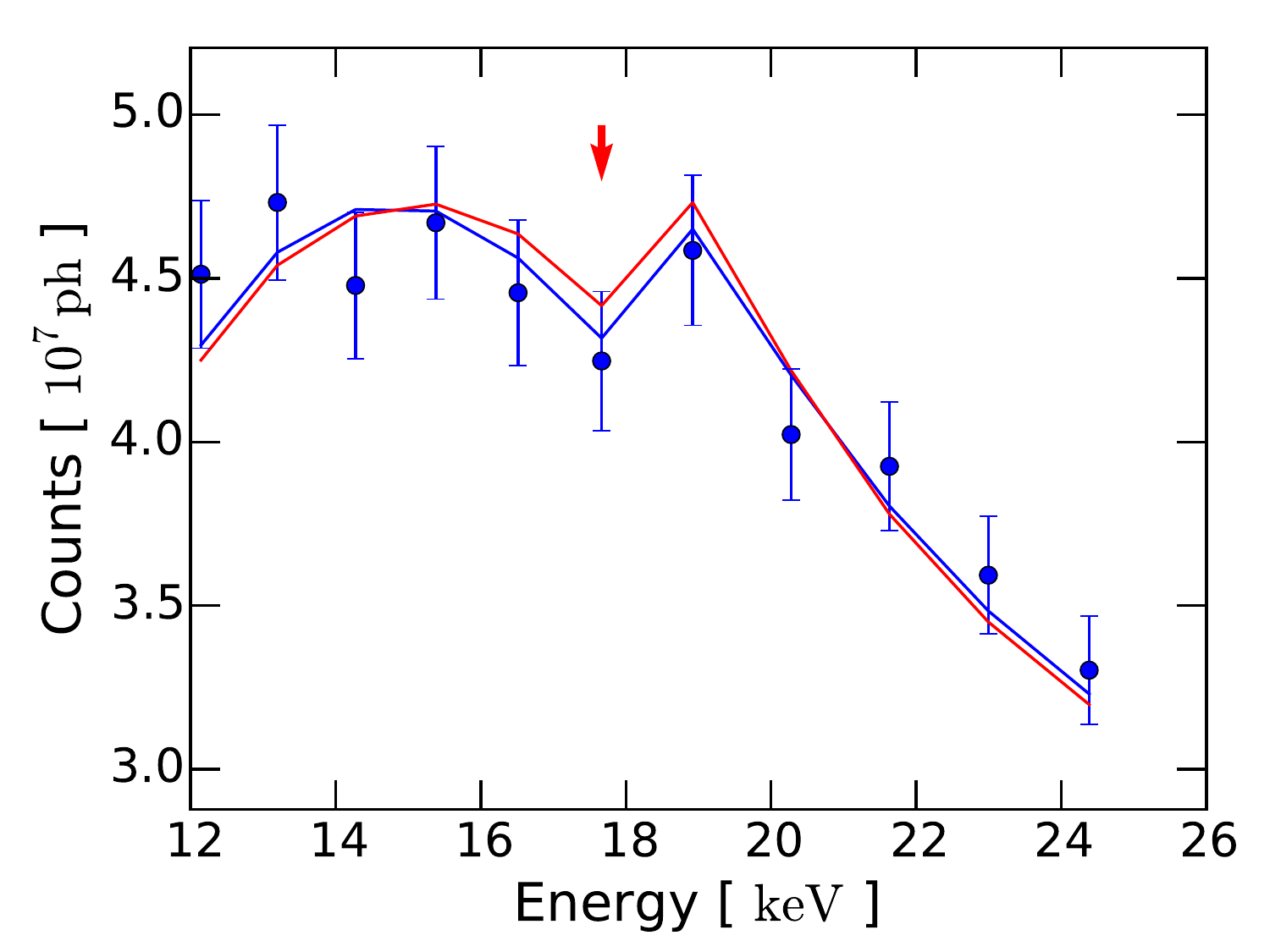}
\includegraphics[angle=0.0,width=0.3\textwidth]{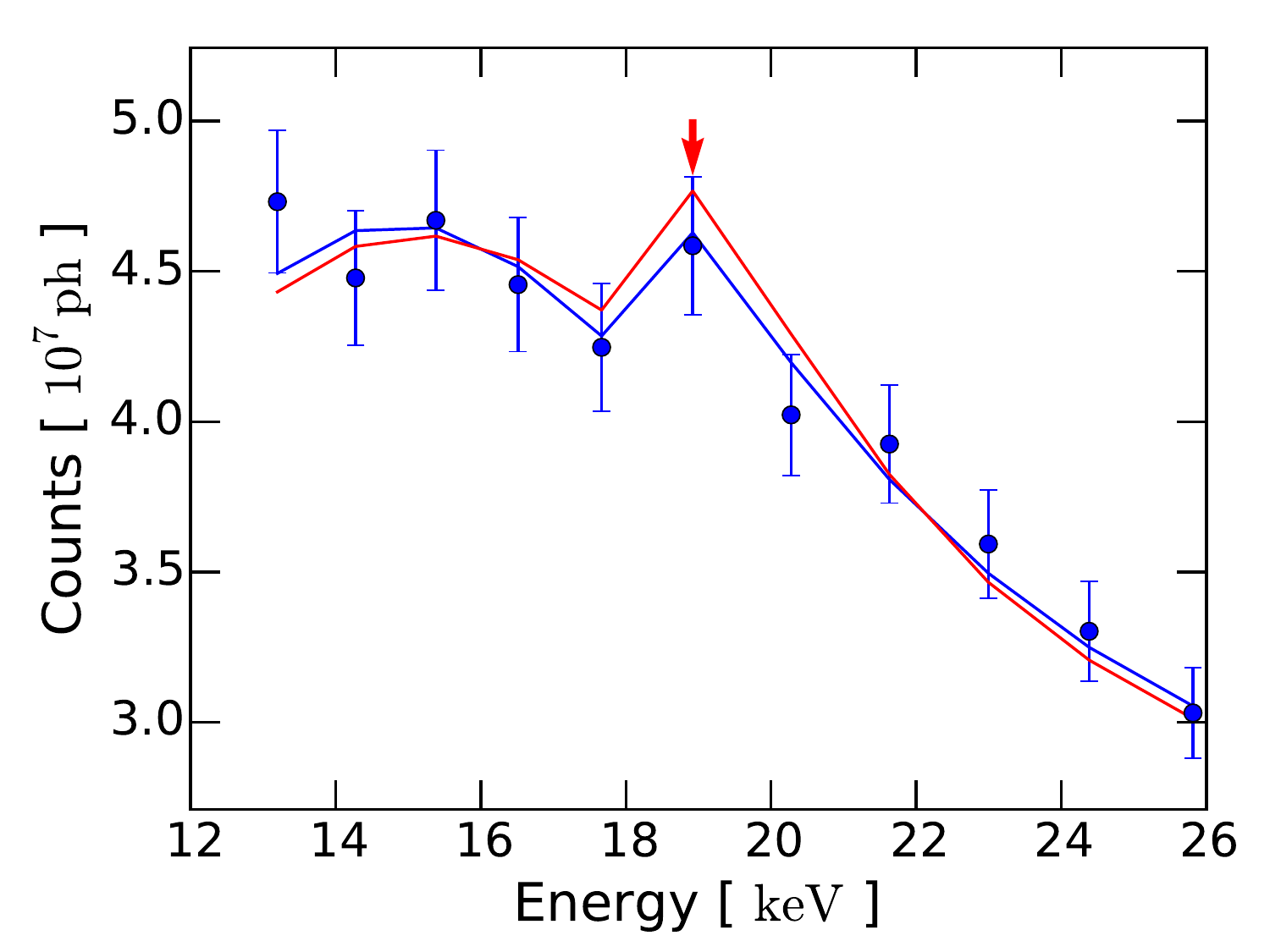}
\includegraphics[angle=0.0,width=0.3\textwidth]{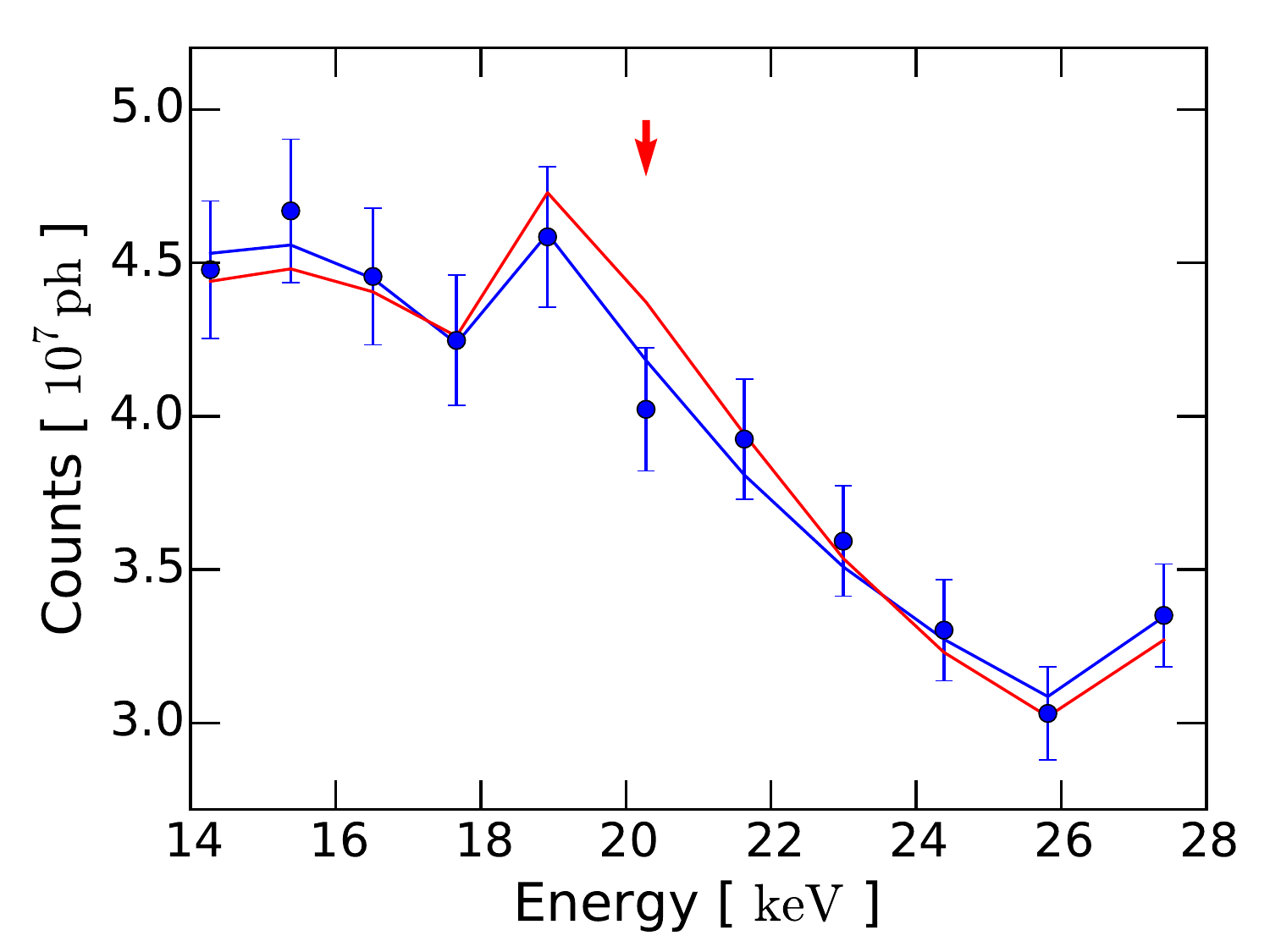}
\includegraphics[angle=0.0,width=0.3\textwidth]{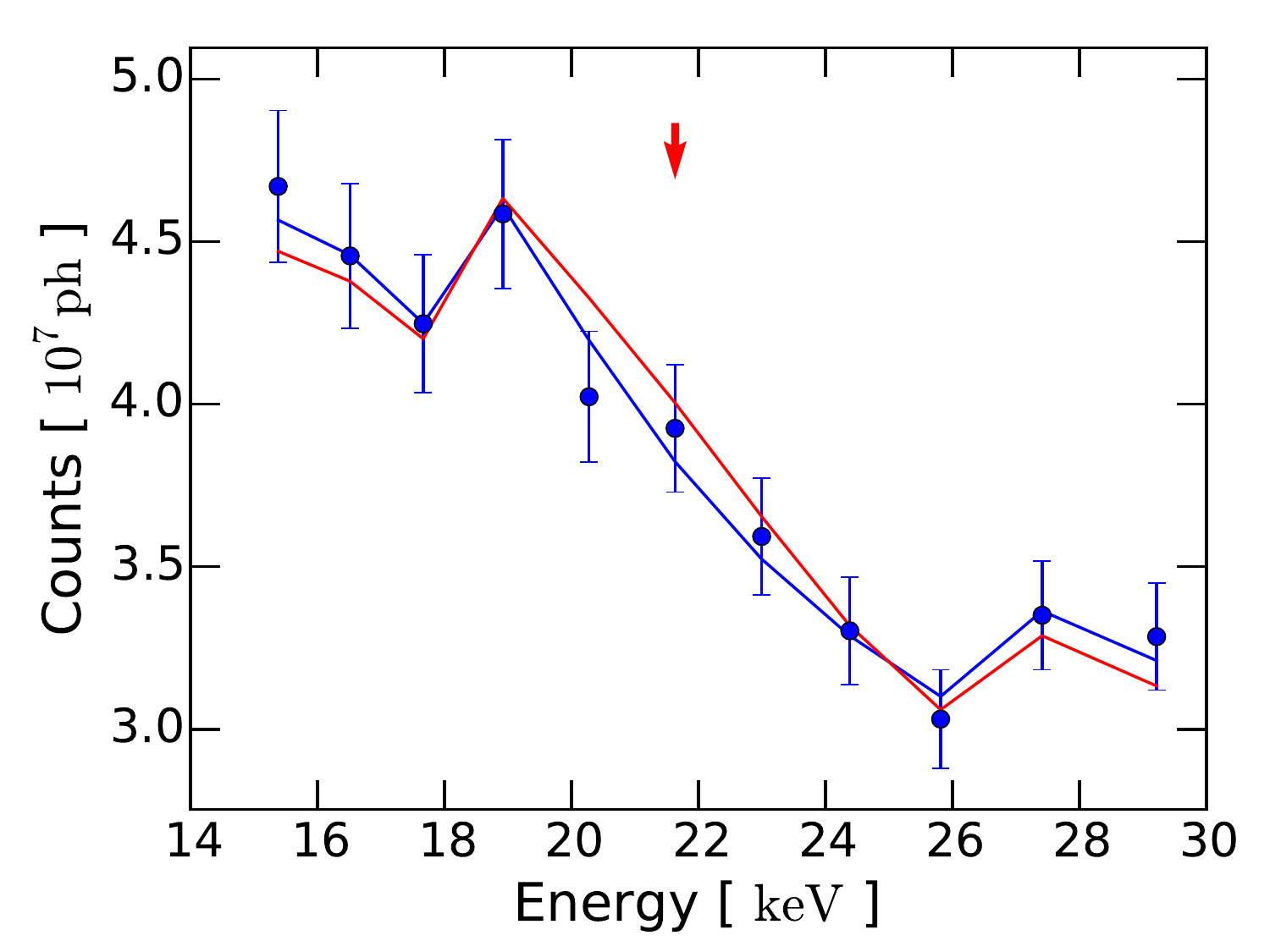}
\includegraphics[angle=0.0,width=0.3\textwidth]{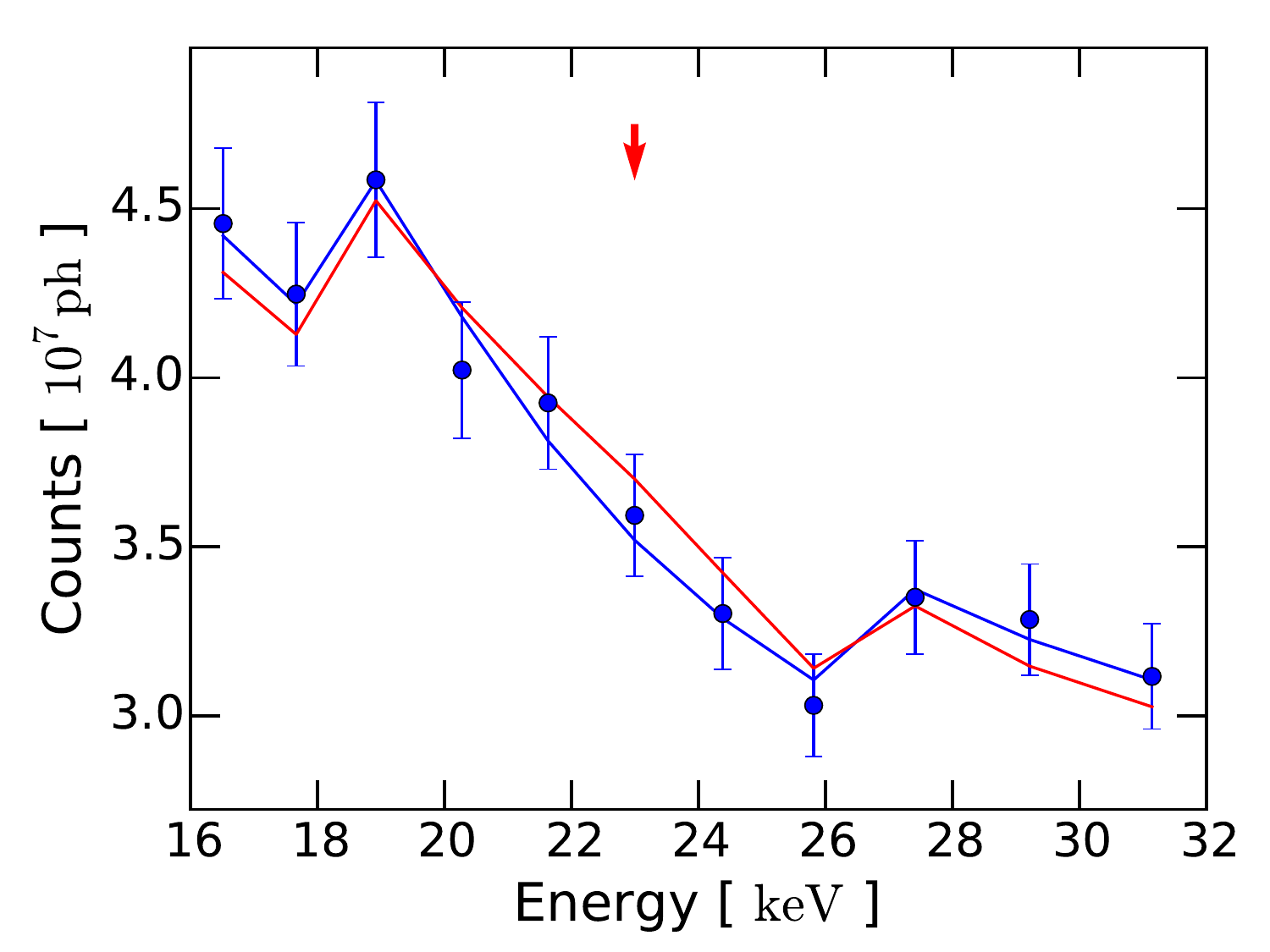}
\includegraphics[angle=0.0,width=0.3\textwidth]{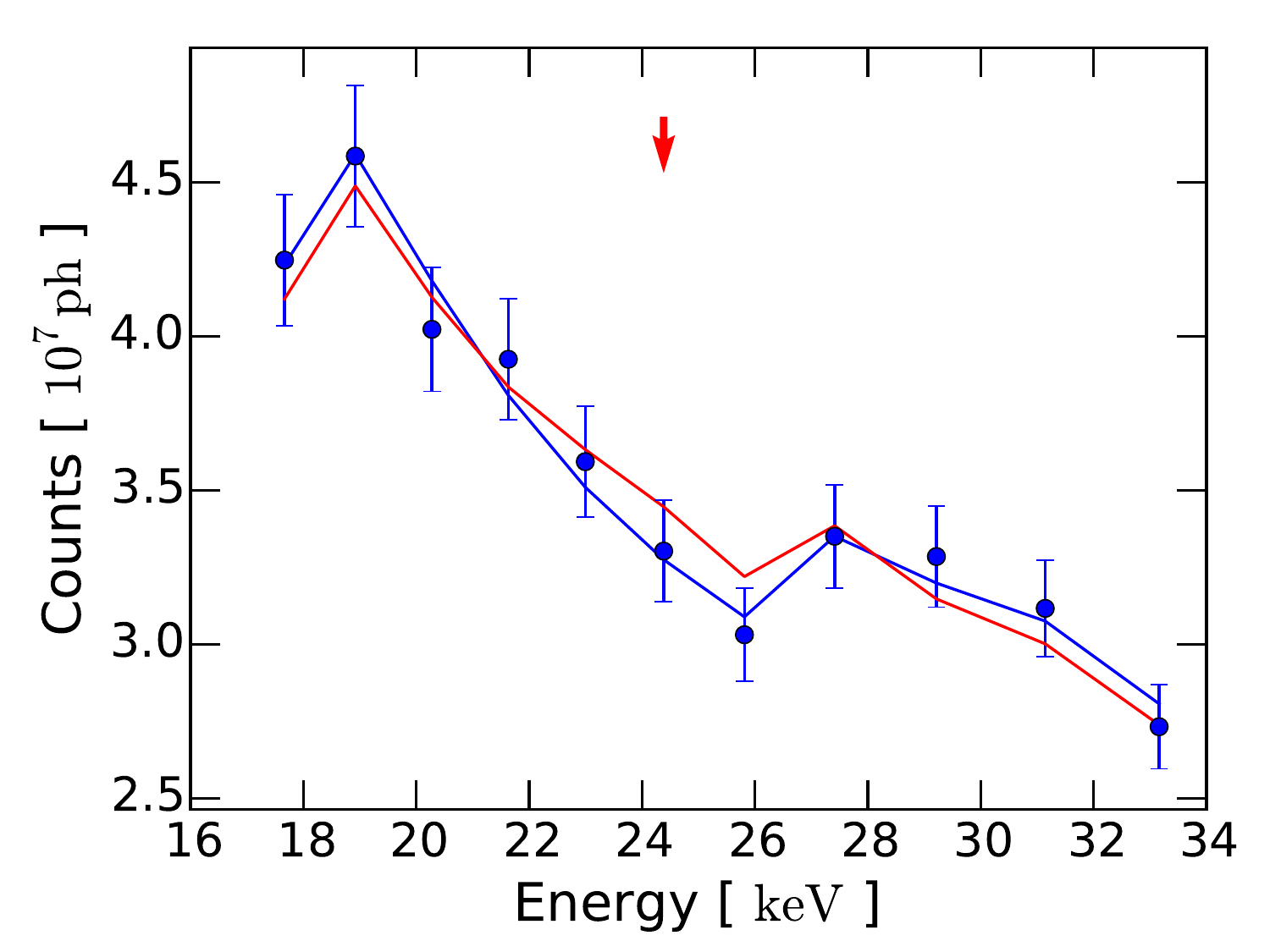}
\includegraphics[angle=0.0,width=0.3\textwidth]{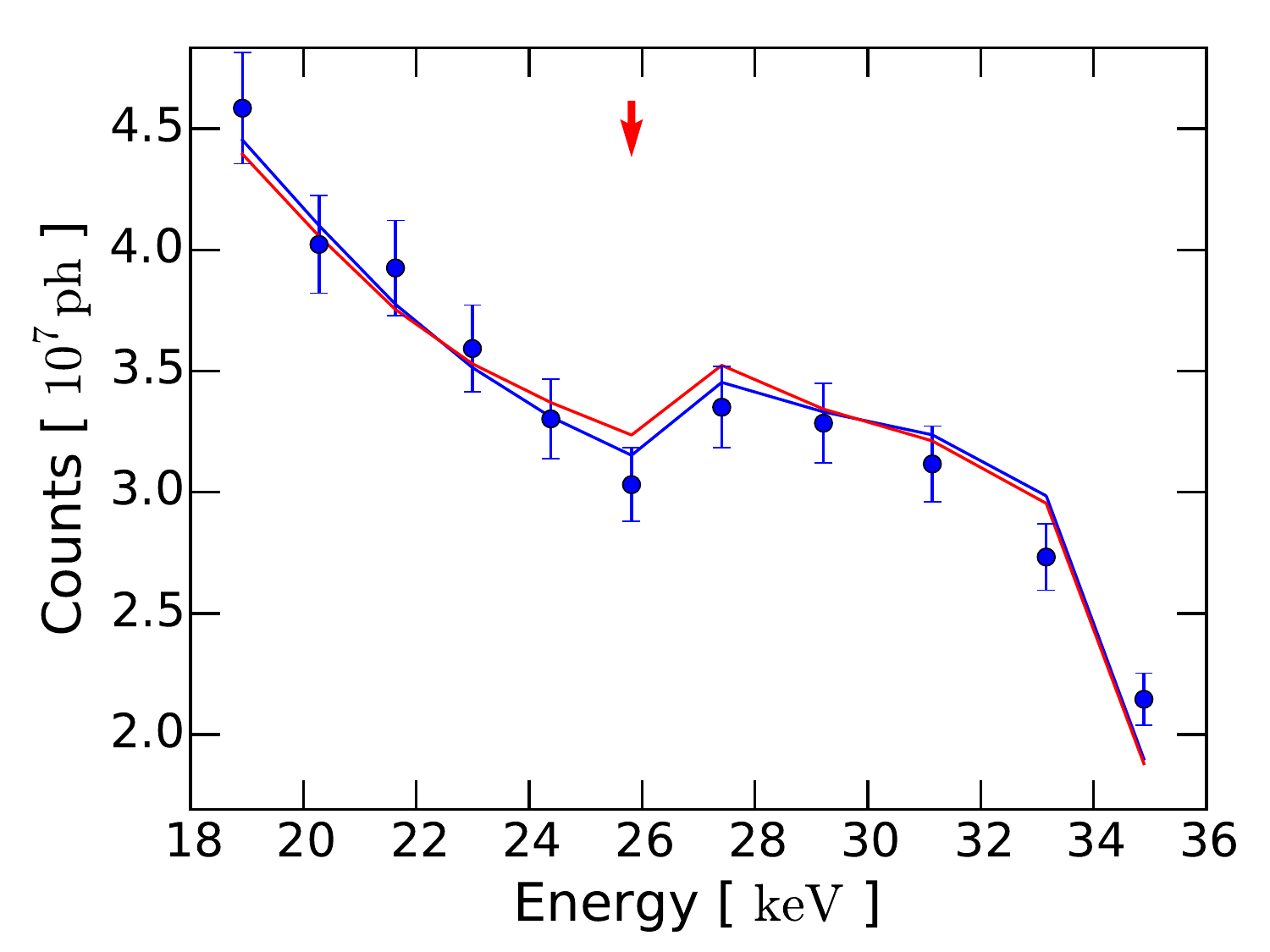}
\caption{The blue points are the measured data with error bars indicating the 5\% systematic error.  The blue line is the best fit model to the data with only the power-law component.  The red line shows the best fit model when including the line signal with 95\% upper limit amplitude.  The red arrow indicates the central energy of the line signal.  }
\label{fig:fit_array}
\end{figure*}

The total expected counts in energy bin $i$ in the search window is then obtained by convolving with the detector energy resolution and integrating the model over the energy bin,  
\begin{equation}\label{eq:data_model}
\nu_{i} + b_{i} = \int_{E^{\rm min}_{i}}^{E^{\rm max}_{i}} dE \int d\tilde{E} \left( \frac{dv}{d\tilde{E}} +\frac{db}{d\tilde{E}} \right)  G(\tilde{E},E) \, .
\end{equation}
Comparing the data model~(Eq.~(\ref{eq:pre_fs})) with the expected signal~(Eq.~\ref{eq:data_model})), the line amplitude $f_{s}$ is related to sterile neutrino parameters by 
\begin{eqnarray}\label{eq:fs_definition}
f_{s} &=& \frac{\rho_\odot R_\odot} {4\pi m_s \tau}\sum T_{j}\tilde{J}(E_{0},j)  \\
&=&8.6 \times 10^{-2} {\rm cm^{-2} \, s^{-1} \, sr^{-1}}    \left(\frac{\sin^{2}2\theta}{10^{-10}} \right) \left(\frac{m_{s}}{10\; {\rm keV}} \right)^{4} \times \nonumber \\
&&\sum T_{j}\tilde{J}(E_{0},j) \nonumber \, .
\end{eqnarray}

To search for a line signal from the data, it is important to understand the uncertainties associated with the measurement.  We first consider the systematic uncertainty in the effective area of the detector, which is $\sim5$\% according to the GBM collaboration \cite{Bissaldi:2008df, Meegan:2009qu}.  Note the quoted uncertainty is the \emph{total} uncertainty for the effective area, which in principle can be two different kinds of uncertainty.  The first kind is the overall uncertainty on the effective area across all energy bins, which affects the value of the flux obtained from data.  The second kind is the uncorrelated errors between energy bins, which may introduce spurious spectral features even if the true flux spectrum and the true effective area are both smooth in energy.  For a spectral analysis, the uncorrelated error among energy bins is much more important than simply a normalization shift.  In this work, we conservatively attribute all the 5\% uncertainty to the uncorrelated errors. As a result, the model uncertainty for each energy bin is
\begin{equation}
\sigma_{\rm Aeff}  = 0.05 (\nu_{i} + b_{i})  \, .
\end{equation}
We then consider the statistical uncertainty.  As shown in Fig.~\ref{fig:counts}, the number of photons is enormous in the energy range that we are interested in.  The statistical uncertainty in each bin is small, $\sqrt{N}/N<10^{-3}$.  Therefore, we safely ignore the statistical uncertainties in this work. 

We adopt the method of maximum likelihood for fitting the counts spectrum for each search window.  For each search bin $i_{0}$, we assume a Gaussian probability distribution function for each energy bin in the search window, giving the likelihood function:
\begin{equation}
{\cal L}(f_{s}, { \Xi} | i_{0}) = \prod_{i} \frac{1}{\sqrt{2 \pi} \sigma_{\rm Aeff} } e^{-\frac{{ (\nu_{i} + b_{i}-d_{i})^{2}}}{2 {\sigma_{\rm Aeff}^{2}} } } \, ,
\end{equation}
where the product is taken over the energy bins $i$ in the search window.  Best fit parameters are obtained by maximizing the likelihood function, or equivalently minimizing its negative logarithm.

\begin{figure}[t] 
\includegraphics[width=3.5in]{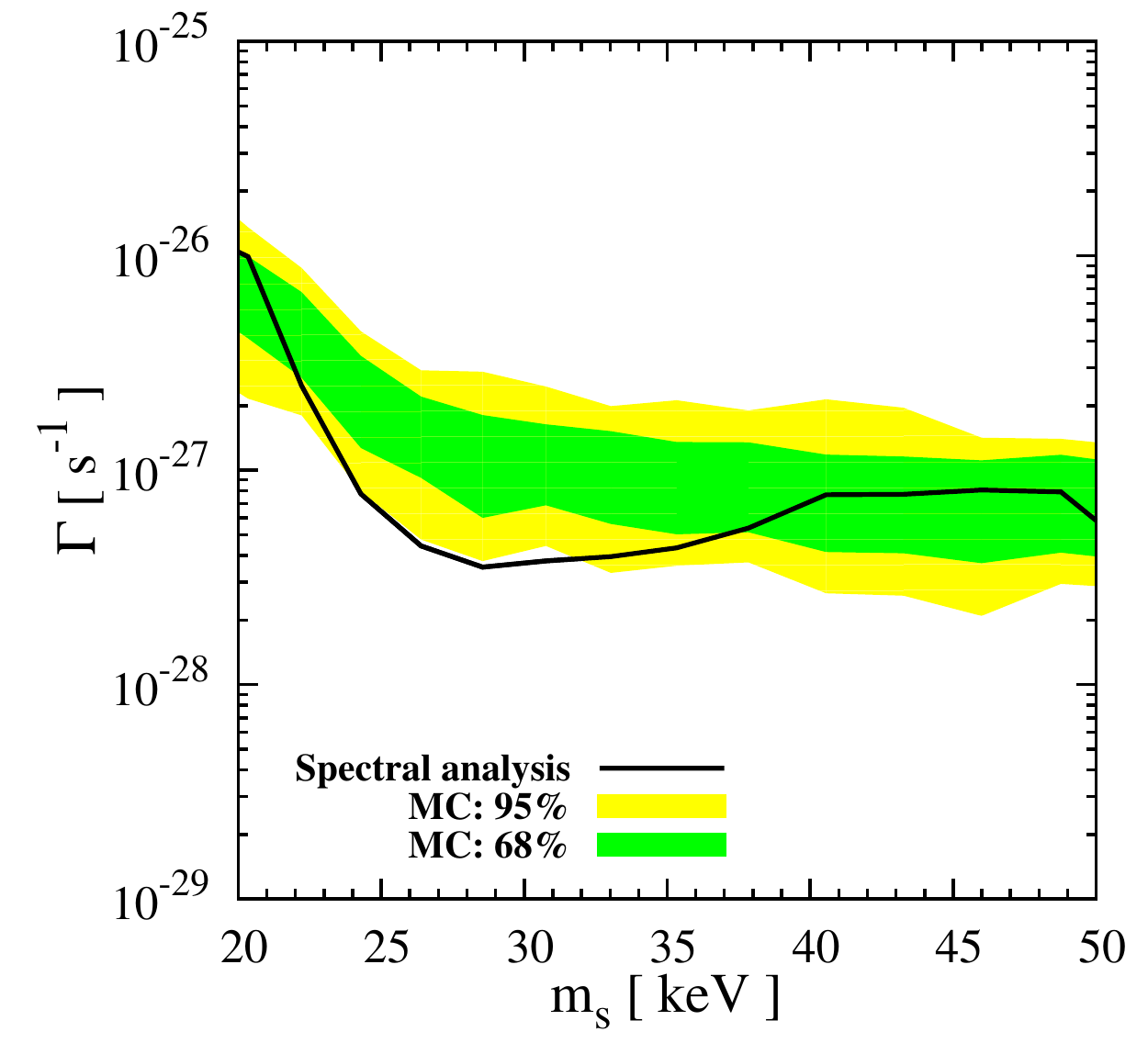} 
\caption{ The black solid line is the 95\% C.L. upper limit from the spectral analysis.  The green~(yellow) shaded region shows the 68\%~(95\%) intervals from the Monte Carlo simulations.  }
\label{fig:gbm_mc}
\end{figure}

We first find the best-fit background only parameters, $\Xi_{0}$, where $f_{s}$ is set to zero.  The implicitly defined $\Xi_{0}$ is given by
\begin{equation}
\lambda(f_{s} = 0,\Xi_{0}|i_{0}) = {\rm Min} \{ -2{\rm Log}{\cal L}(f_{s}=0, {\Xi}\, | i_{0}) ; \Xi  \} \, .
\end{equation} 
We check whether the power-law only background model is a reasonable hypothesis by computing the reduced $\chi^{2}$~($\chi^{2}$ per degree of freedom) for each search window.  We find that the reduced $\chi^{2}$ ranges from 0.2 to 1.4 in our analysis range.  We therefore conclude that the power-law only model plus the prescribed 5\% systematic error can reasonably describe the data for each search window.  

In Fig.~\ref{fig:fit_array}, we show explicitly the 15 search windows for this analysis.  The blue data points are the GBM data, and the assigned error bars are the 5\% systematic uncertainty.  The statistical errors are too small to be shown.  The blue lines are the best fit count spectrum from the power-law only model described above.  The apparent peculiar spectral features, such as those around 18 and 26\,keV, are successfully captured by the power-law model when effective area and non-uniform energy bins are taken into account.

\subsubsection{Limits on dark matter decay rate \label{sec:gammalimit}}

To search or constrain the line signal, we use the so-called profile likelihood method~\cite{Rolke:2004mj}. We search for the best fit line amplitude by minimizing the negative log-likelihood with respect to all the model parameters, 
\begin{equation}
{\lambda}(f_{s0}, \Xi_{0}|i_{0}) = {\rm Min} \{ -2{\rm Log} {\cal L}(f_{s}, { \Xi} | i_{0}) ; f_{s}, \Xi  \} \, , 
\end{equation}
where $f_{s}$ is constraint to be non-negative. 
We observe no significant preference for the presence of the line signal.  We then proceed to find the 95\% C.L.~one-sided upper limits on the dark matter signal amplitude, $f_{s}^{95}$, by increasing the amplitude while continuously minimizing the log-likelihood function over the nuisance parameters, until it is 2.71 larger than the best fit log-likelihood, 
\begin{equation}
 {\rm Min} \{ -2{\rm Log} {\cal L}(f_{s}^{95}, { \Xi} | i_{0}) ;  \Xi  \} \equiv {\lambda}(f_{s 0}, \Xi_{0}|i_{0})+2.71\, .
\end{equation}
The best fit model when the line signal is at 95\% upper limit is shown in Fig.~\ref{fig:fit_array} in red lines.  The red arrow indicates the energy of the inserted X-ray line.  Using $f_{s}^{95}$ and Eq.~(\ref{eq:fs_definition}), we then obtain the 95\% C.L.~upper limit of the dark matter decay rate.  

\begin{figure}[t] 
\includegraphics[width=3.5in]{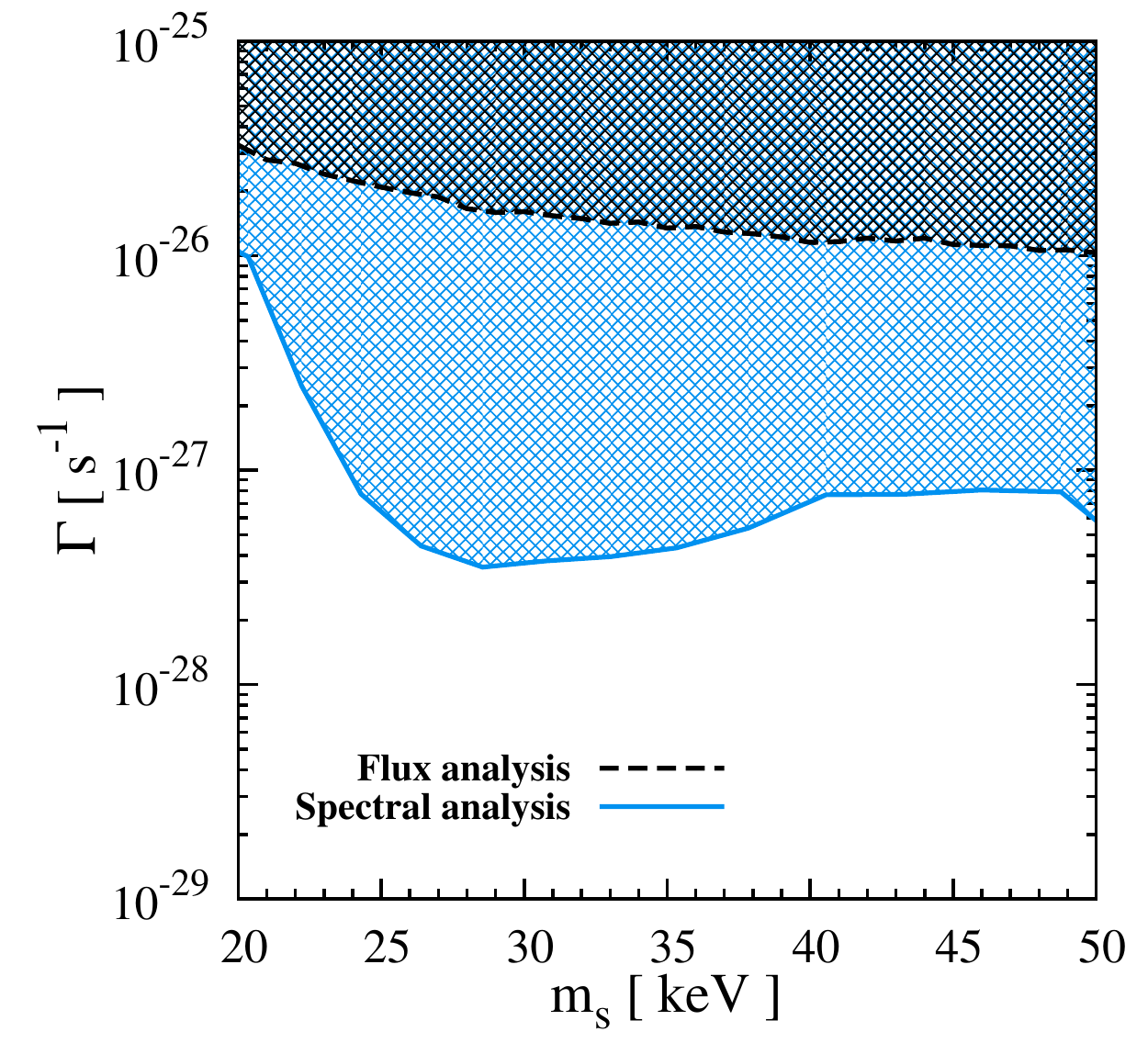} 
\caption{The conservative upper limit from the flux analysis~(Black) and the 95\% C.L.~upper limit from the spectral analysis~(Blue) for the decay rate of sterile neutrino dark matter.  The hashed regions are excluded by the corresponding analyses.  }
\label{fig:gammalimit}
\end{figure}

We perform a Monte Carlo study to check the robustness of the limit.  For each search window, we generate 100 mock data sets, according to the best fit power-law only parameters and the 5\% systematic error with Gaussian probability distribution function.  We perform the profile likelihood analysis to obtain the 95\% upper limits for the mock data sets.  In Fig.~\ref{fig:gbm_mc}, we show the obtained upper limit from data and the 68\% and 95\% coverage of the limits from our Monte Carlo simulations.  

Overall, we find that the observed limit is consistent with our Monte Carlo realizations at the 95\% level.  At about 28\,keV sterile neutrino mass, we find the actual limit touches the 95\% lower bound of the expect limit.  This is likely due to the data point at about 14\,keV falls below what one would expect from a smooth power-law flux spectrum, as shown in the fits in Fig.~\ref{fig:fit_array}.  There are no known detector defects at this energy~\cite{Bissaldi:2008df, Meegan:2009qu}, we thus consider this as a $\sim 2\sigma$ systematic downward fluctuation in the effective area model.  This downward shift effectively means the data prefers a negative line, which results in the improved limit at this energy.  It is also important to note that the limits are correlated due to the largely overlapping data points in adjacent search windows.  As a result, limits from line energies close to 14\,keV are all slightly improved.  The simulated limits from our Monte Carlo realizations are not correlated with adjacent line energies, since the mock data sets are generated independently for each search window.

Finally, Fig.~\ref{fig:gammalimit} shows the limits obtained on the decay rate from both the flux analysis and the spectral analysis, with the hashed region corresponding to the excluded parameter space.  As expected, the spectral analysis produces a much stronger limit than the flux analysis.  Since the presence of a line signal mostly only affects one energy bin, one would expect the spectral analysis limit is approximately given by the size of the error bars of the data points, and thus the spectral analysis limit is expected to be about 5\% of the flux analysis limit.  This is indeed the case in most of the mass range, except where the data prefers a negative line, as discussed above.  At low energies, the spectral analysis limit deteriorates rapidly.  This is due to the imposed lower cutoff of the search energy window.  As the line energy approaches the boundary, the number of bins used for the fit is reduced accordingly, and the spectral shape becomes increasingly degenerate with the power-law shape.  Both factors cause the limit from spectral analysis to deteriorate.

\subsection{Limits on Sterile Neutrino Dark Matter \label{sec:snulimit}}

Using the upper limits on the decay rate, we derive the corresponding upper limit on the mixing angles for sterile neutrino dark matter.  In Fig.~\ref{fig:constraint}, we show the constraint on the mixing angle--mass plane.  For comparison, we also show the only limit in this energy range, obtained with CXB observations using \emph{HEAO-1}~\cite{Boyarsky:2006fg}, and with Milky Way observations using \emph{INTEGRAL}~\cite{Boyarsky:2007ge}.  Unsurprisingly, the flux analysis does not yield competitive limits. However, in the mass range  $m_{s} \sim$ 25 -- 50\,keV, the spectral analysis improves the limit on the mixing angles by about an order of magnitude compared to the previously strongest limit.  

Compared to the previous analysis~\cite{Boyarsky:2006fg}, which uses \emph{HEAO-1} A4 Low Energy Detector data~\cite{Gruber:1999yr}, our analysis improves mainly in:  The GBM data has smaller error bars compared to the \emph{HEAO-1} data~($\sim 5\%$ vs $\sim 10\%$);  We have employed several cuts to reduce cosmic rays induced backgrounds;  The Milky Way halo yields a larger signal flux than the CXB alone;  The GBM NaI detector has a slightly better energy resolution.  These factors, each expected to give a factor of few improvement, all contribute to our improved limit.

\section{Discussion and conclusion\label{sec:disconclu}}

\subsection{Future Developments}
\begin{figure}[t] 
\includegraphics[width=3.5in]{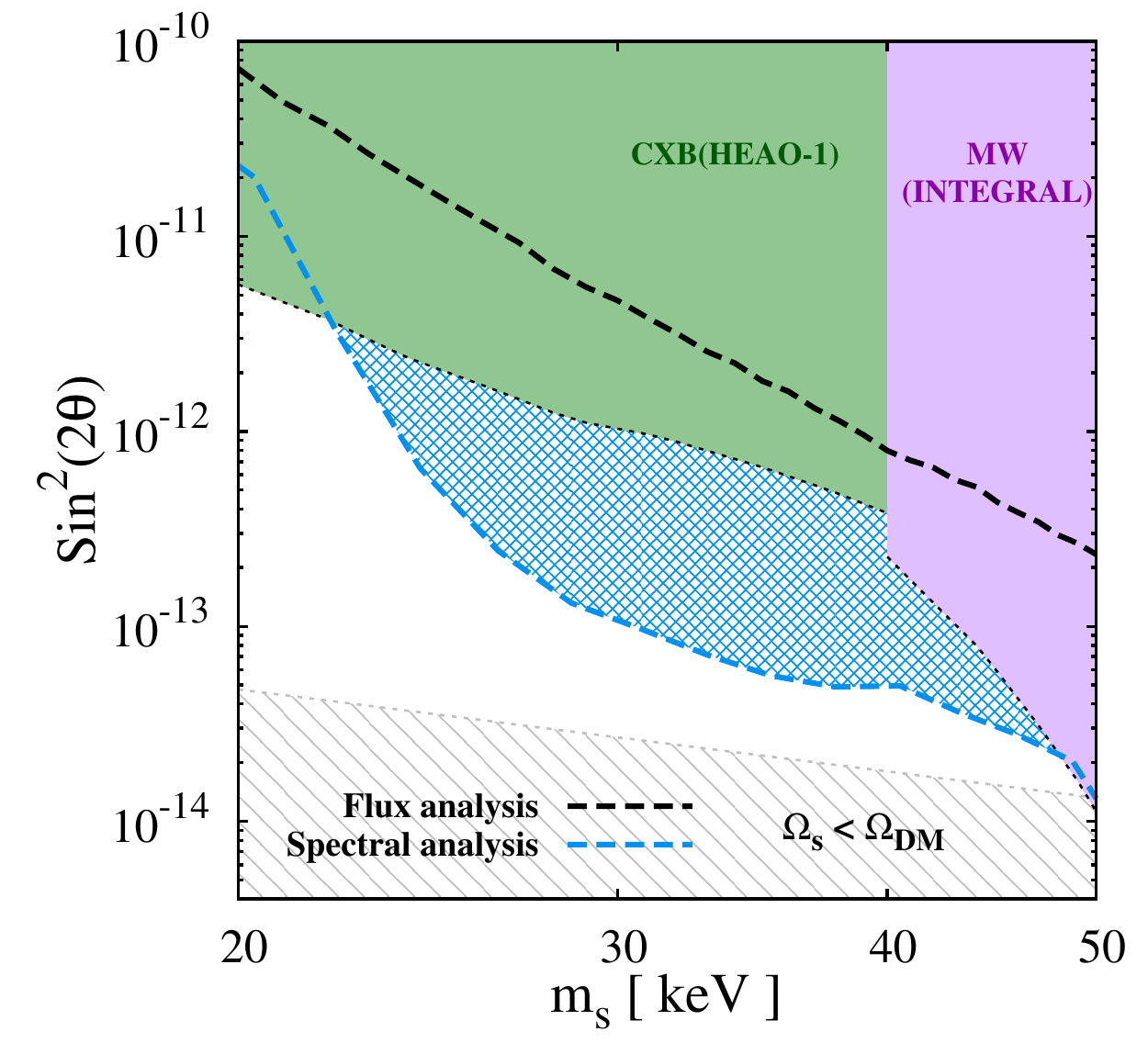} 
\caption{ Constraints from X-ray missions on sterile neutrino dark matter decays, which depends on the mixing angle, $\sin^{2}(2\theta)$, and the mass, $m_{s}$.  In the shown mass range, the best previous constraints are set by observations of cosmic X-ray background (CXB) from \emph{HEAO-1}~\cite{Boyarsky:2006fg} and the Milky Way (MW) halo from \emph{INTEGRAL}~\cite{Boyarsky:2007ge}. The lower bound of the parameter space\,(black-hashed region) is valid for the model $\nu$MSM~\cite{Boyarsky:2009ix}.  The flux~(spectral) analysis limit derived from this work is shown in black-dashed~(blue) line.
  }
\label{fig:constraint}
\end{figure}

In this work, we obtain competitive limits on sterile neutrino dark matter decay by analyzing GBM data. This is the first time the GBM data is used for a dark matter search, and we have obtained the strongest constraint available in the mass range $25-50$\,keV~(Fig.~\ref{fig:constraint}). Although we focus on the implications for the sterile neutrino, our limits can be applied to all dark matter candidates that produce a mono-energetic photon in the keV range, such as moduli dark matter, gravitino dark matter, and other candidates~\cite{Kusenko:2012ch, Essig:2013goa, Albert:2014hwa}.  It is straightforward to constrain the parameter space for the corresponding dark matter candidates using the limit on the decay rate from Fig.~\ref{fig:gammalimit}, taking into account any normalization or energy scaling.  

Our analysis uses simple data reduction, minimal background assumptions, and straightforward analysis procedures.  Thus there are many ways our results  can be improved with further study. 

Firstly, the dominant uncertainty on the data currently comes from the energy behavior of the effective area.  In this work, we use a constant 5\%  uncertainty as quoted by the GBM collaboration, which we then validate against a power-law assumption.  The limit can be improved if this uncertainty can be better quantified or calibrated, e.g., using known background lines.  

Secondly, most of the observed counts come from cosmic-ray related events.  In this work we do not attempt to model such background in detail.  In principle, this background can be better understood with simulations of cosmic-ray interactions with the satellite that take into account the satellite geometry and composition.  Additionally, one can characterize the cosmic-ray induced background by satellite positioning, either by using high energy observations where the data is background dominated, or using the Earth's magnetic field information.  One can possibly construct a template for cosmic-ray induced events, which allows background reduction using spatial information.  A significant reduction of the cosmic-ray background can further improve the dark matter limit for all energies. 

The next source of backgrounds for dark matter searches is those arising from astrophysical origins, which dominate at low energies.  This background can be modeled using high resolution X-ray sky maps from other missions.  One can generate an astrophysical template for the GBM detectors which can then be used to subtract the astrophysical contribution from the data.  

Importantly, if eventually the data are reduced to a regime where statistical uncertainties become important, it is important to treat systematic and statistical uncertainty simultaneously~\cite{Albert:2014hwa}.  

In principle, the analysis can be extended to higher energies.  The GBM-NaI detectors are sensitive up to 1~MeV and the GBM-BGO detectors extend to 40\,MeV, which is even higher than {\it INTEGRAL} and complementary to \emph{ COMPTEL}.  However, at higher energies, the NaI detectors start to observe photons from backward directions due to either re-scattering or penetrating photons.  The BGO detectors are designed to be sensitive to both front and back directions.  A dedicated analysis taking into account this detector response for signal modeling is therefore necessary.  

\subsection{Astrophysical Implications}
It is important to highlight that our novel use of GBM successfully detects the Galactic astrophysical component at 10--20\.keV energies, as shown in Fig.~\ref{fig:counts_rate_map} and~\ref{fig:counts}.  Due to the broad point spread function of the NaI detector, this seemingly diffuse astrophysical component contains all points source emission near the GC, and possibly some diffuse emission.  GBM observations may be used to impose interesting limits on the total Galactic astrophysical emission, which would extend the results from INTEGRAL~\cite{2005ApJ...635.1103B, 2008ApJ...679.1315B, 2011ApJ...739...29B} down to $\sim 10$\,keV.  

To constrain the astrophysical component, it is necessary to further reduce the detector background~(see suggestions in the previous section).  The analysis procedures would also need to be modified for a continuum spectrum.  Such an analysis and a detailed interpretation of the astrophysical component are beyond the scope of this work.

\subsection{Conclusion}
We use data obtained by a GBM NaI detector\,(det-7), on board \emph{Fermi} to set limits on dark matter decaying into mono-energetic photons.  We first perform a conservative flux analysis, based on comparing the total flux normalization of the data and the model.  We then perform a spectral analysis that assumes the total of non-dark matter contributions exhibits a power-law flux spectrum within the search window.  Our spectral analysis is able to improve the limit by about an order of magnitude compared to previous searches using CXB with {\it HEAO-1} data for line energies in the 10--25\,keV energy range, or 20--50\,keV sterile neutrino mass.  

Conventionally used to detect and locate transients such as gamma-ray bursts, this is the first time that the GBM has been used to construct an all-sky map and search for dark matter emissions.  After performing careful background reduction procedures, we are able to detect the astrophysical component centered on the GC.  

Dark matter searches with GBM benefit from the large sky coverage and good energy resolution. Although the GBM does not have excellent angular resolution, this is not a severe problem for decaying dark matter for which the expected Milky Way signal is appreciably extended.  A unique advantage of the GBM-NaI detectors is that they are sensitive to an energy range that is too high for X-ray telescopes such as {\it Chandra}, {\it Suzaku}, and {\it XMM-Newton}, but too low for {\it INTEGRAL}, therefore filling an energy gap that was last probed by {\it HEAO-1} in the late 1970s. 

The current search sensitivity is dominated by systematic uncertainties in the effective area.  A better understanding of GBM detector response through simulation or calibration, as well as a better understanding of detector and astrophysical backgrounds, can further improve the data quality and resulting limits substantially.

\section*{Acknowledgments}
We especially thank John Beacom for many helpful comments throughout the course of this work.  We thank Shirley Li for helpful comments and lending CPU power, and Mark Finger (Universities Space Research Association Huntsville) for providing transient cuts for the GBM data.  This work is supported by NASA grant NNX11AO46G.  K.C.Y.N. was supported by NSF Grant PHY-1101216 and PHY-1404311 to John Beacom.  S.H. was supported by a Research Fellowship for Research Abroad by JSPS.  J.G. acknowledges support from NASA through Einstein Postdoctoral Fellowship grant PF1-120089 awarded by the Chandra X-ray Center, which is operated by the Smithsonian Astrophysical Observatory for NASA under contract NAS8-03060. 

\bibliographystyle{h-physrev}
\bibliography{bibliography}

\end{document}